\documentclass[proof]{pasj01}

\Received{$\langle$reception date$\rangle$}
\Accepted{$\langle$acception date$\rangle$}
\Published{$\langle$publication date$\rangle$}
\usepackage{times}
\usepackage{url}
\usepackage{lineno}

\begin{document}

\title{Orbital modulations of X-ray light curves of Cyg X-1 in its low/hard and high/soft states}
\author{Juri Sugimoto$^{1,2}$, Shunji Kitamoto$^{2,3}$, Tatehiro Mihara$^1$, Masaru Matsuoka$^1$\\}
\altaffiltext{}{
$^1$  MAXI team, Institute of Physical and Chemical Research (RIKEN), 2-1 Hirosawa, Wako, Saitama 351-0198, Japan\\
$^2$  Department of Physics, College of Science, Rikkyo University, 3-34-1 Nishi-Ikebukuro, Toshima, Tokyo 171-8501, Japan\\
$^3$  Research Center for Measurement in Advanced Science,\\
 Rikkyo University, 3-34-1 Nishi-Ikebukuro, Toshima, Tokyo 171-8501, Japan
}
\email{sugimoto@crab.riken.jp}

\maketitle

\begin{abstract}

The black hole binary Cygnus X-1 has a 5.6 day orbital period.
We first detected a clear intensity modulation 
with the orbital period in its high/soft state with 6 year MAXI data, 
as well as in its low/hard state.
In the low/hard state, 
the folded light curves showed an intensity drop at the superior conjunction of the black hole 
by a modulation factor ($MF$), which is the amplitude divided by the average,
with $8\pm1 \%$, $4\pm 1 \%$ and $3\pm 2 \%$  for  $2-4$ keV, $4-10$ and $10-20$ keV bands,
showing a spectral hardening at the superior conjunction of the black hole.
Spectral analysis in the low/hard state, 
with a model consisting of a power law and a photoelectric absorption, 
showed that 
the hydrogen column density, $N_{\rm H}$, increased 
from $(2.9 \pm 0.4) \times 10^{21}\ {\rm cm}^{-2}$ 
to $(4.7 \pm 1.1) \times 10^{21}\ {\rm cm}^{-2}$
around the superior conjunction. 
The flux of the power law component decreased with  $6\pm 1\%$.
On the other hand, 
$MF$s of the folded light curves in the high/soft state, 
were $4\pm 1\%$ and $4\pm 2\%$ for $2-4$ keV and $4-10$ keV bands, respectively.
We applied a model consisting of a power law and a diskblackbody with a photoelectric absorption.
A modulation of the flux of the power law component was found with $7\pm 5 \%$ in $MF$,  
while the modulation of   $N_{\rm H}$ was less than $1\times 10^{21}\ {\rm cm}^{-2}$.
These results can be interpreted as follows;  
the modulation of both states can be mainly explained by scattering of X-rays by an ionized stellar wind, 
but, only at the superior conjunction in the low/hard state, a large photoelectric absorption appears 
because of a low ionization state of the wind in the line of sight at the phase 0.
Such a condition can be established by reasonable parameters of an in-homogeneous wind and the observed luminosities.
\end{abstract}

\section{Introduction}

The black hole (hereafter BH) binary Cygnus X-1 (hereafter Cyg X-1) is persistently bright in the X-ray band. 
It shows typically two spectral states \citep{oda, remillard2006, done}, 
the low/hard state (hereafter LHS) 
which is dominated by a power law spectrum with a high energy cutoff, 
and the high/soft state (hereafter HSS) which is dominated 
by optically-thick thermal emission from an accretion disk, 
i.e. the standard-disk \citep{shakura, tanaka, remillard2006, done}. 
Cyg X-1 repeats transitions between the two states in several days to several hundred days.
The companion, HDE 2268968, is a  O9Iab super giant star.
An orbital period of the binary is  5.6 days \citep{mason, bolton} 
and the inclination angle is 27$^{\circ}$ \citep{orosz}.
A strong stellar wind from the companion star is captured by the  black hole 
and probably forms a characteristic structure around the black hole like a focused wind \citep{friend}.
The wind exposed by luminous X-rays should be ionized and also considered to be clumpy \citep{church, feng}. 

Although an eclipse is not seen in the X-ray light curve, 
an orbital modulation in the X-ray intensity has been observed 
in its LHS \citep{priedhorsky, wen, kitamoto2000, misko}, 
especially in the low energy band. 
However, the intensity modulation has not been detected 
in the HSS \citep{wen, brocksoppA}.
\citet{boroson} reported that 
a peak of 5.61 d in the periodogram of the hardness ratio was detected 
during a part of the HSS period, 
using the RXTE/ASM data, 
but it was weaker than that in the LHS.

The absorption dip was first detected in the LHS \citep{li}, 
which is an abrupt intensity drop in the soft X-ray band 
around the superior conjunction of the BH.
Various depth and the duration of dips have been observed \citep{kitamoto1984, church}. 
The duration of the dip is distributed from several seconds to more than 10 min.
The probability of the dip is high around superior conjunction, 
at phase $0.9-0.1$.
The cause of the dips is interpreted by an absorption 
by clumpy and in-homogeneous stellar wind \citep{kitamoto1984, boroson, misko}.
\citet{yamada} discovered a dip during the HSS by a Suzaku/XIS observation. 
They detected 
Fe absorption edge at 7.5 keV 
and He-like and H-like Fe$-$K$_{\alpha}$ absorption lines 
in the energy spectrum around the dip, 
indicating absorption by a highly ionized wind.

In order to clarify the wind condition and its difference between the two states,
we analyzed more than 6 yr observational data with Monitor of All sky X-ray Image (MAXI) \citep{matsuoka} 
and compared the orbital modulation in the LHS and the HSS.

\section{Observation}
\label{sec:obs}

MAXI is attached to the International Space Station.
As the International Space Station orbits the earth in every 92 min, 
MAXI scans over nearly the entire sky with two kinds of X-ray cameras: 
the Gas Slit Camera (GSC: \cite{mihara}) covering the energy band of $2-20$ keV,
and the Solid-state Slit Camera (SSC: \cite{tomida}) covering $0.7-7$ keV.
The long term variation of Cyg X-1 using the same data has been reported by \citet{sugimoto2016} (hereafter Paper I).
The data extraction-criteria and observation details are described in Paper I.

\begin{figure*}
  \begin{center}
   \includegraphics[width=15cm]{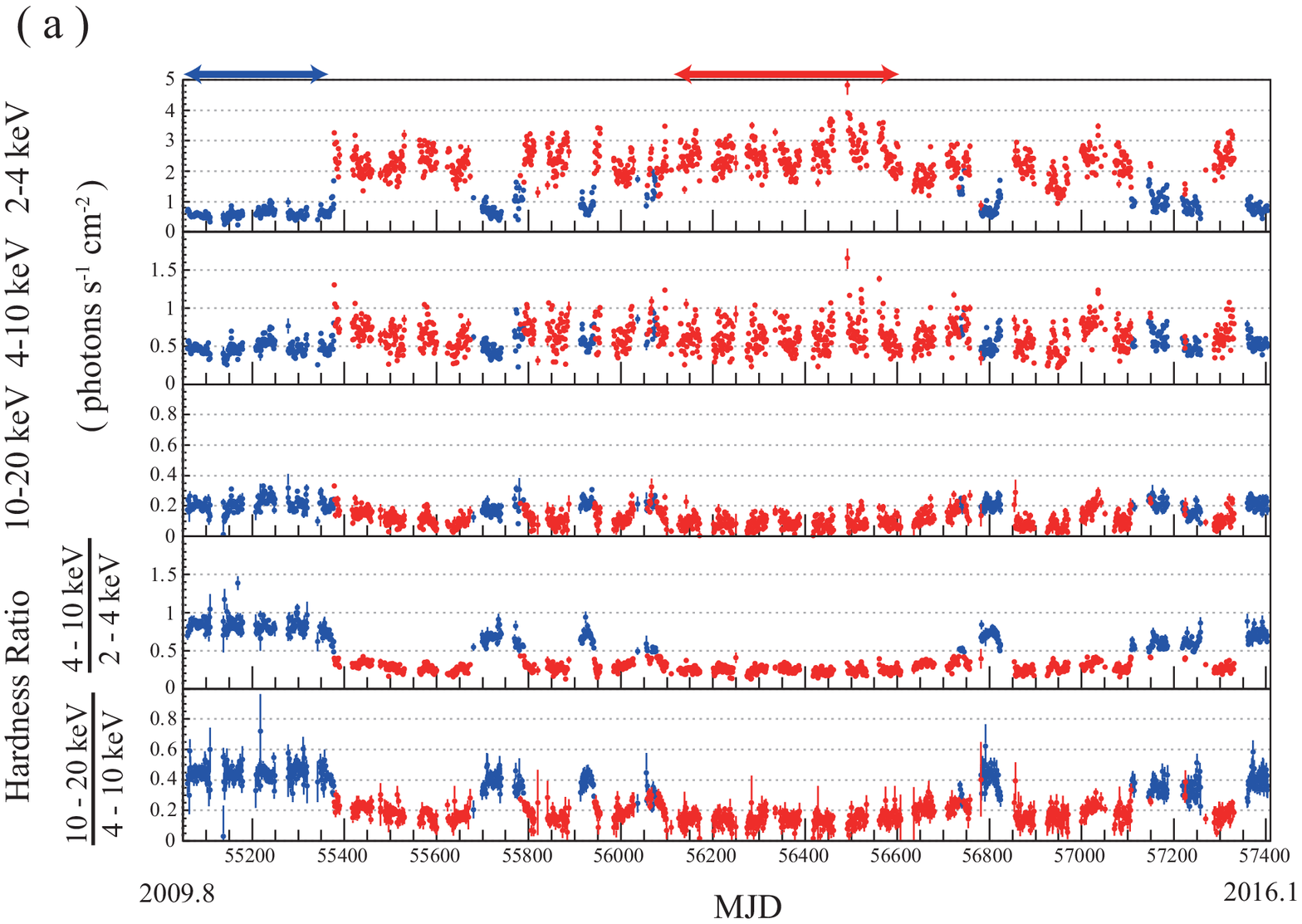}\\
   \includegraphics[width=7cm]{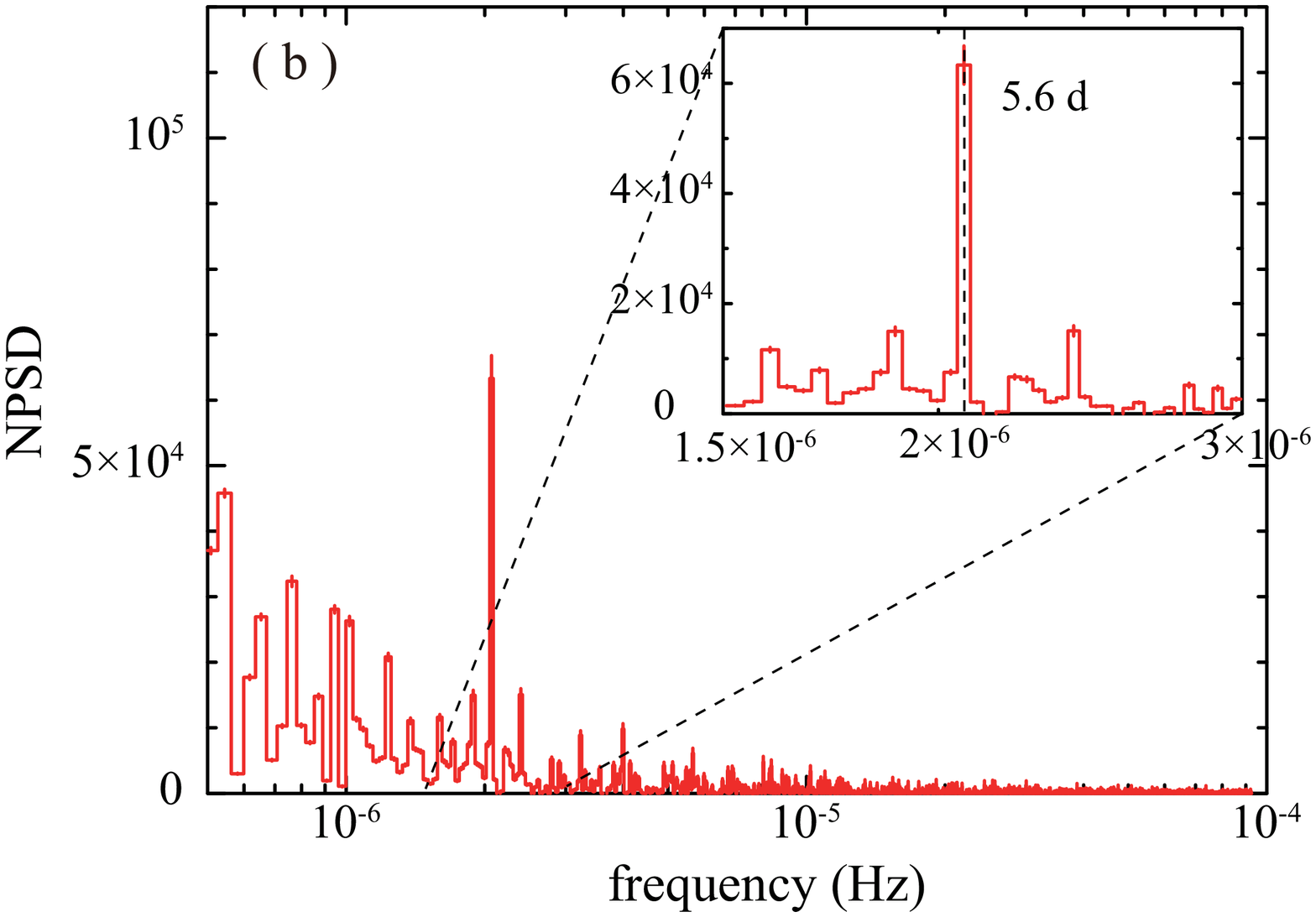}
   \hspace{1cm}
   \includegraphics[width=7cm]{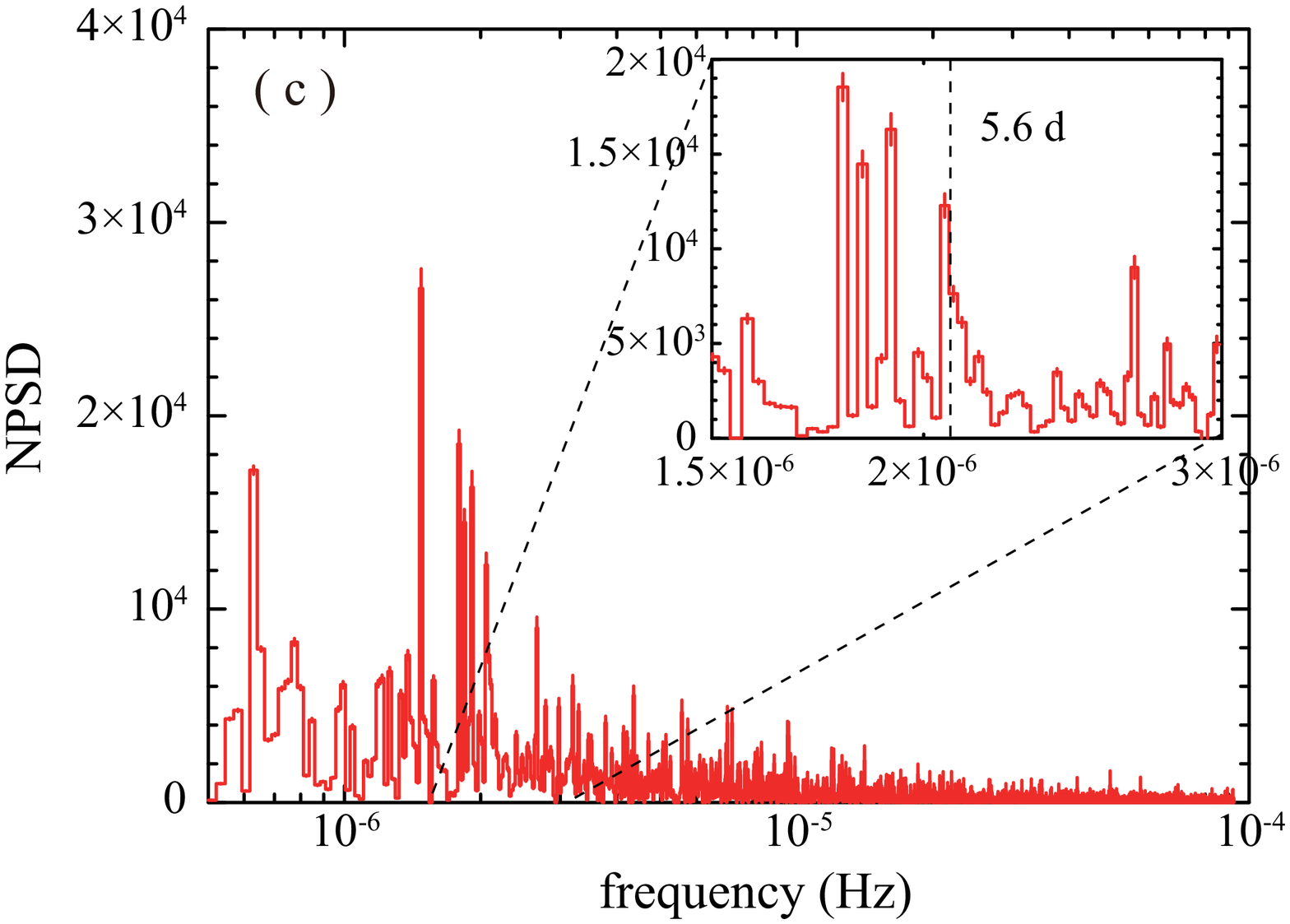}
  \end{center}
  \caption{(a) One-day bin light curves and HR histories of Cyg X-1 obtained with the MAXI/GSC.
  From the top to bottom panels,  $2-4$ keV, $4-10$ keV and $10-20$ keV intensities,
  and  $I$($4-10$ keV)/$I$($2-4$ keV) and $I$($10-20$ keV)/$I$($4-10$ keV) ratios are plotted.
  The blue and red points show the LHS and the HSS periods, respectively.
  The arrows indicated the data span for the calculation of the power spectrum.
  (b)(c) The power spectra in the $2-4$ keV band in the LHS and the HSS, respectively.}
 \label{fig:cygLC}
\end{figure*}

Figure \ref{fig:cygLC} (a) shows 
one-day bin light curves of Cyg X-1 obtained with the GSC 
from 2009 August 15 (MJD = 55058) to 2014 November 9 (MJD = 56970),
in three energy bands ($2-4$ keV, $4-10$ keV and $10-20$ keV).
Time histories of two kinds of hardness ratios (HR), 
$I$($4-10$ keV)/$I$($2-4$ keV) and $I$($10-20$ keV)/$I$($4-10$ keV), 
are also plotted.
The HSS and the LHS are indicated with red and blue, respectively,
by following the definition given in Paper I.	

The power spectra in $2-4$ keV in the LHS and the HSS are shown in figure \ref{fig:cygLC} (b) and (c), 
whose data spans for the calculations were 55058-55376(MJD) and 56130-56607(MJD), respectively.
The peak of 5.6 d orbital period is clearly seen in the LHS, 
but, in the HSS, 
the signal is weak and splits into plural peaks (see Paper I for detail).
Some peaks in the HSS, which are larger than that of 5.6d, are instrumental fakes.

\section{Analysis}
\subsection{Folded light curves and hardness ratios}

\begin{figure*}
  \begin{center}
   \includegraphics[width=7cm]{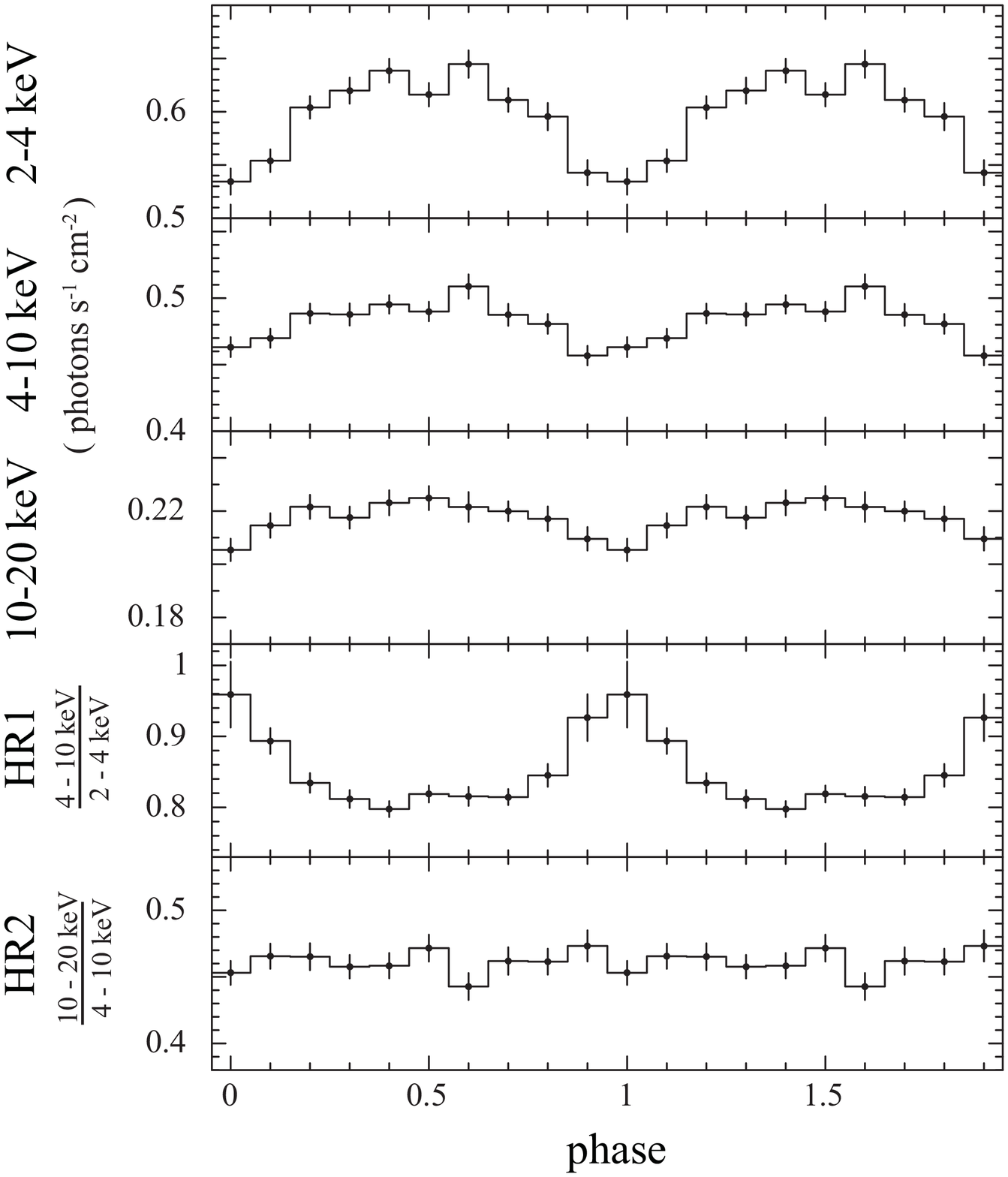}
   \hspace{1cm}
   \includegraphics[width=7cm]{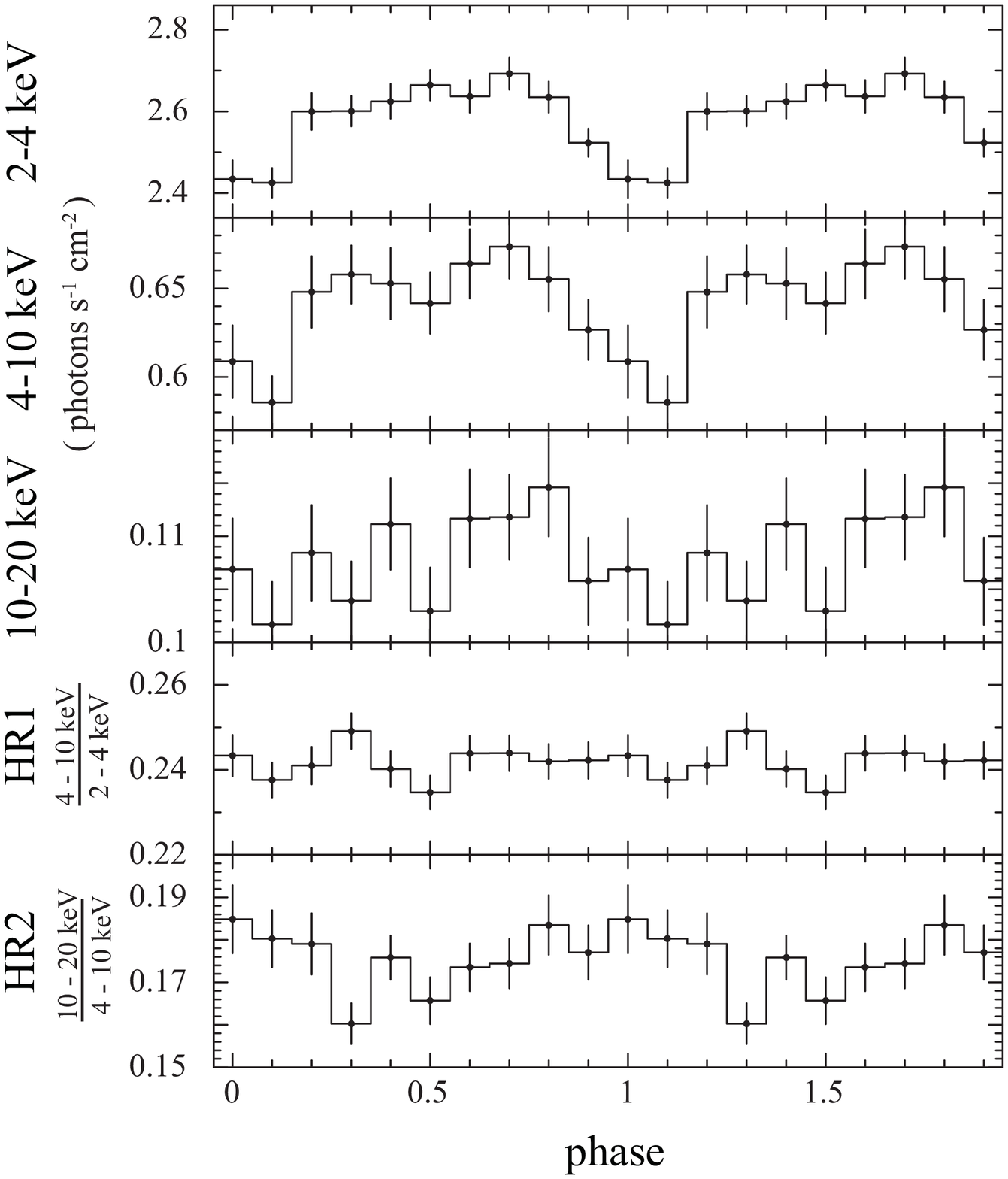}
  \end{center}
  \caption{Folded light curves and HRs in the LHS (left) and the HSS (right).}
\label{fig:foldedlc}
\end{figure*}

We use the orbital period, $P_{\rm orb} = 5.599829\pm 0.000016$ d, 
and an epoch of the inferior conjunction of the O-star, $T_0 = 41874.207 \pm 0.009$MJD, 
which were obtained by \citet{brocksoppB}.
Even if $P_{\rm orb}$ has the maximum error of 0.000016d, 
the difference of the orbital phase at the MAXI observation term is roughly $0.7\%$ 
that is much smaller 
than the phase bin of 0.1 in our analyses.
The folded light curves and HRs are shown in figure \ref{fig:foldedlc}.
The data spans are the same to those of the power spectral analysis.
The errors correspond to standard deviations of the data in each phase.
In the both states, the light curves, 
except for the $10-20$ keV band in the HSS, 
show clear modulation with a minimum around phase 0,  
which corresponds to a superior conjunction of the BH. 
It can be recognized that the modulation amplitude of the $2-4$ keV band in the LHS is larger than others.
To evaluate the modulation quantitatively, 
we fitted sine function to the folded light curves 
and obtained  the amplitude and the average.
Then, we defined  a modulation factor $MF$ as a ratio of the amplitude to the average.
$MF$s of the three energy bands, in the LHS, 
are $8\pm 1 \%$ in $2-4$ keV, $4\pm 1 \%$ in $4-10$ and $3\pm 2 \%$ in $10-20$ keV, respectively.
Whereas those in the HSS are $4\pm 1 \%$ in $2-4$ keV and $4\pm 2 \%$ in $4-10$ keV.
The data points in the $10-20$ keV band in the HSS show large scattering 
and we can find only an upper limit of $MF$ of $4 \%$.

The HR1, $I$($4-10$ keV)/$I$($2-4$ keV), 
in the LHS shows clear hardening around phase 0, 
whereas it is not seen in the HSS.
The HR2, $I$($10-20$ keV)/$I$($4-10$ keV), 
does not show any significant modulation in the both states.

\subsection{Spectral ratio}

\begin{figure*}
  \begin{center}
   \includegraphics[width=7cm]{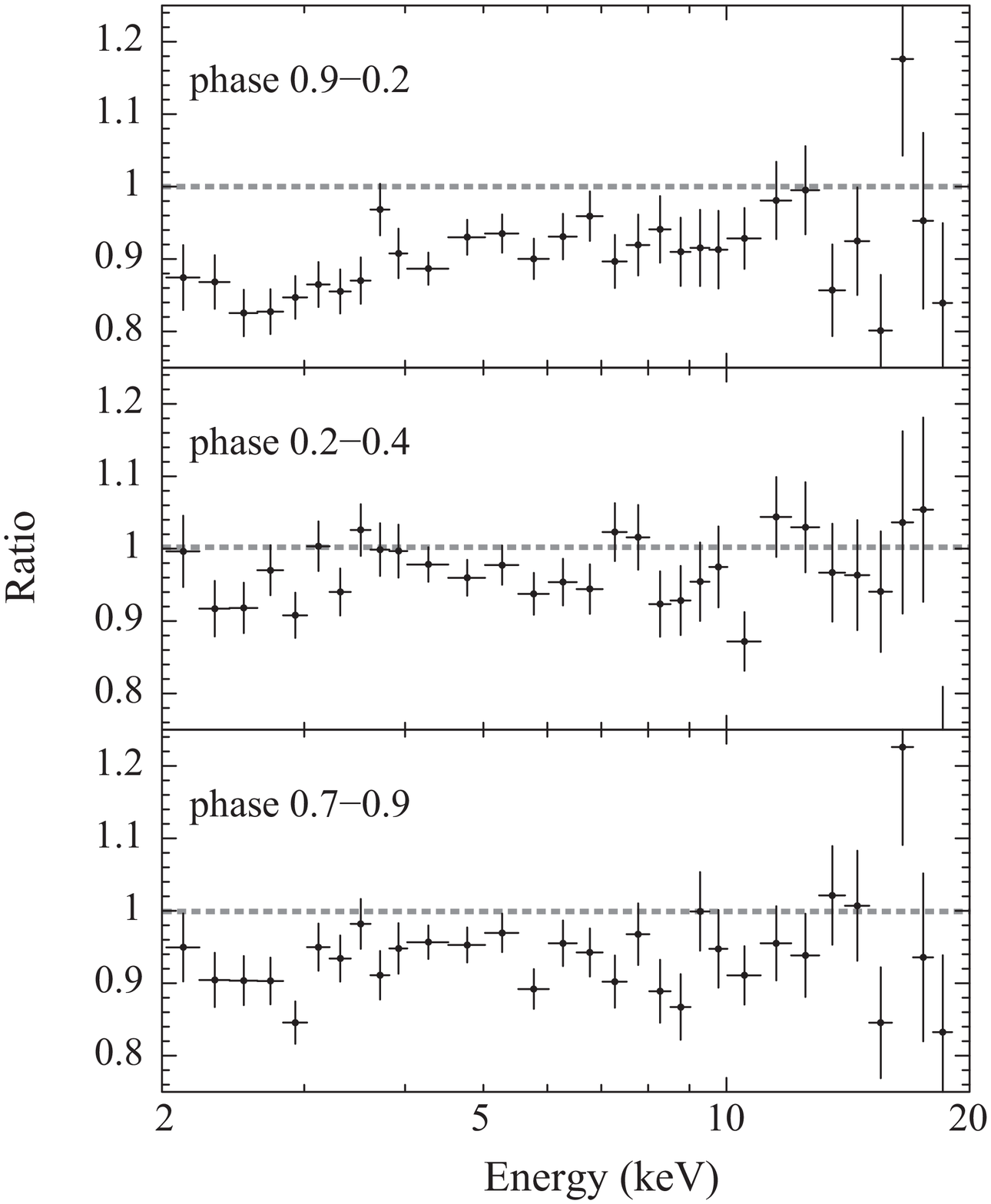}
   \hspace{1cm}
   \includegraphics[width=7cm]{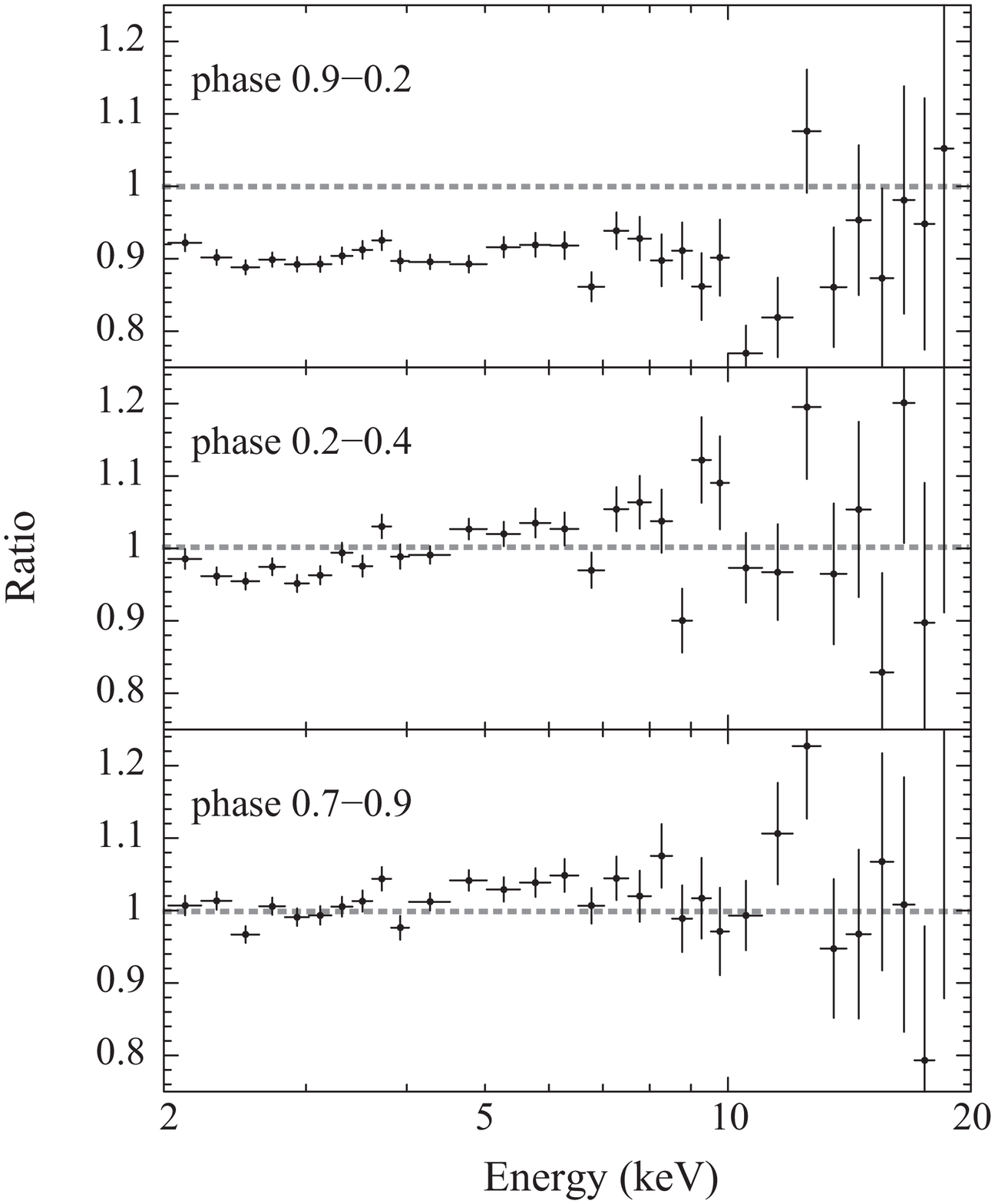}
  \end{center}
  \caption{ Spectral ratios in the LHS (left) and the HSS (right).
  The spectra in the three phases, 0.9-0.2, 0.2-0.4 and 0.7-0.9 were divided by that in the phase 0.4-0.7, 
  for both states.
  }
\label{fig:spec_ratio}
\end{figure*}

We divided the orbital phase into four 
and extracted the energy spectra of the MAXI/GSC in each phase. 
The ratios of the spectrum in each phase to that of the spectrum 
in the phase between 0.4 and 0.7 are presented in figure \ref{fig:spec_ratio}.
In the LHS, 
the ratio of the phase $0.9-0.2$ clearly shows a
significant decrease in the low energy band below 4 keV, 
suggesting an increase of the photoelectric absorption.
However, spectra in the other phases in the LHS and all the phases in the HSS 
do not show an enhancement of the modulation in the low energy band, 
and rather show  an energy independent modulation.

\subsection{Phase resolved spectrum analysis}

\begin{figure*}
  \begin{center}
   \includegraphics[width=7cm]{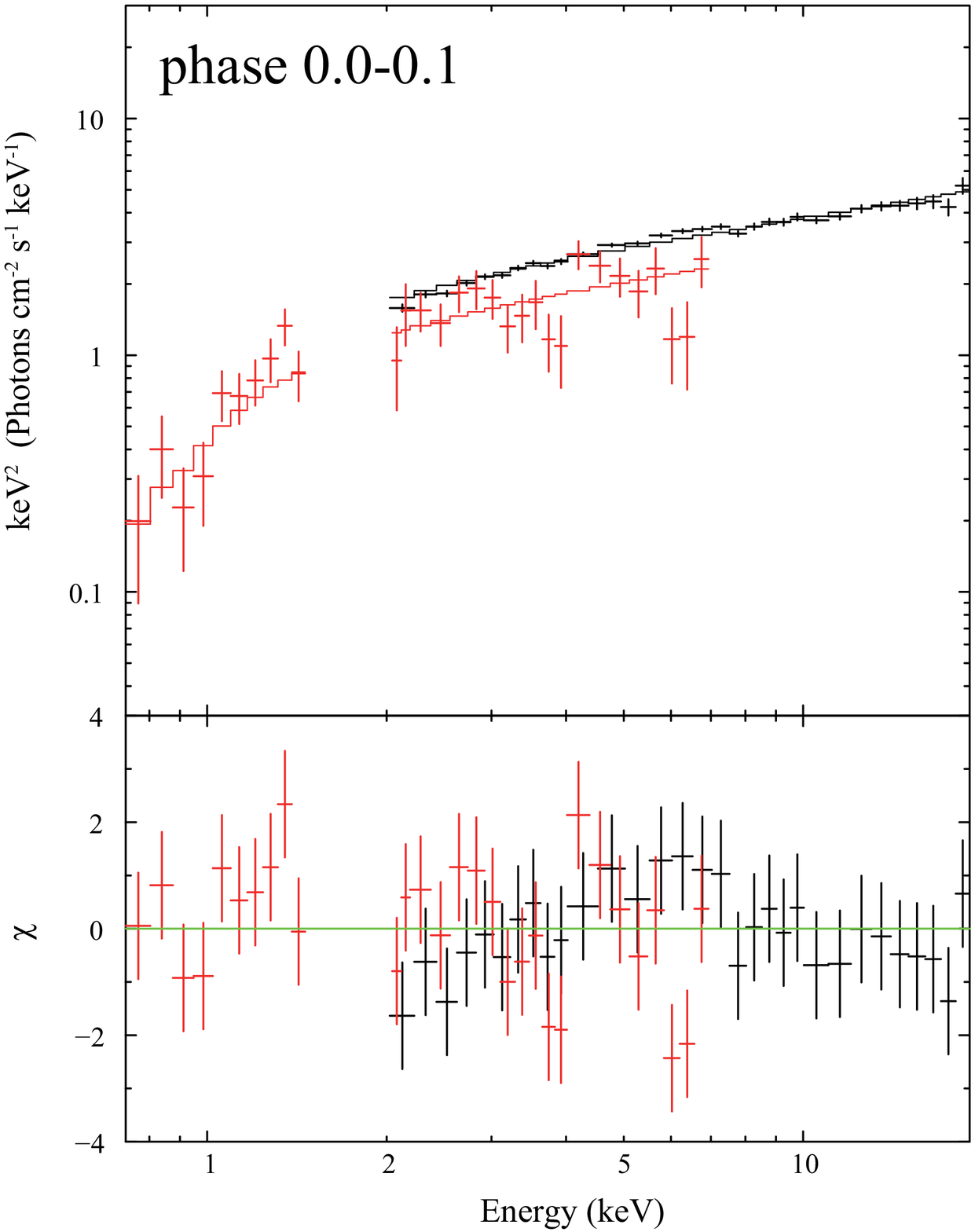}
   \hspace{1cm}
   \includegraphics[width=7cm]{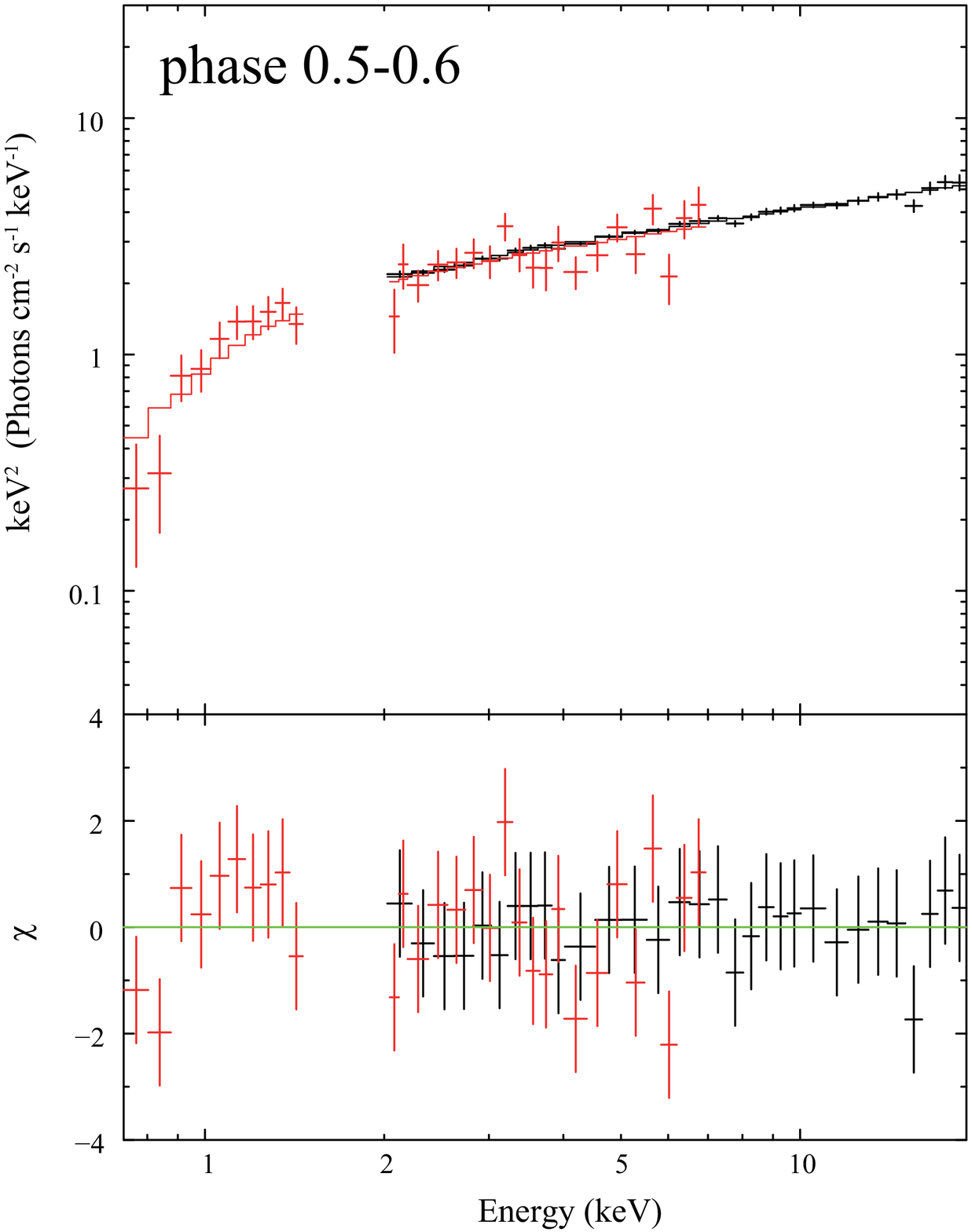}\\
   \vspace{1cm}
   \includegraphics[width=7cm]{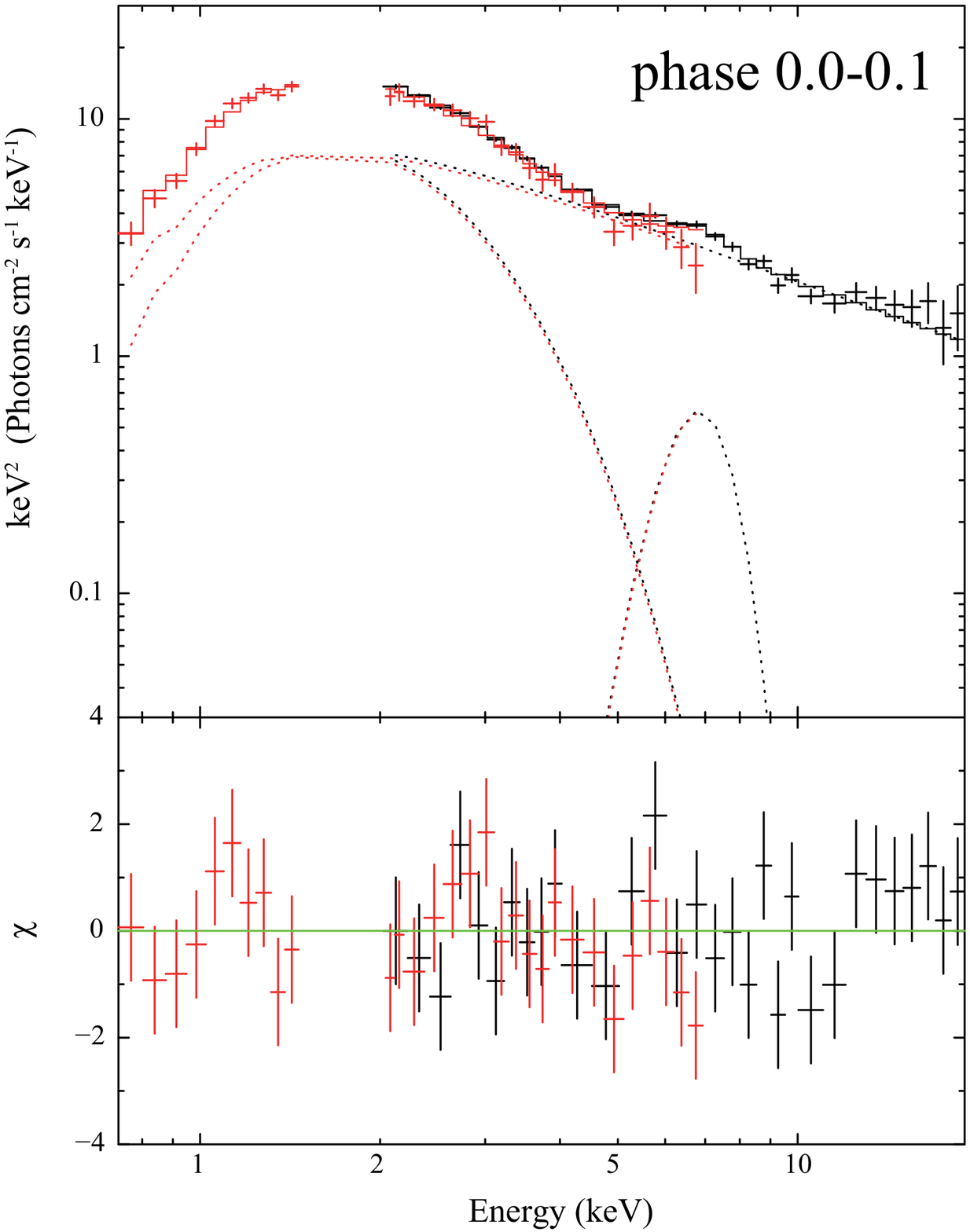}
   \hspace{1cm}
   \includegraphics[width=7cm]{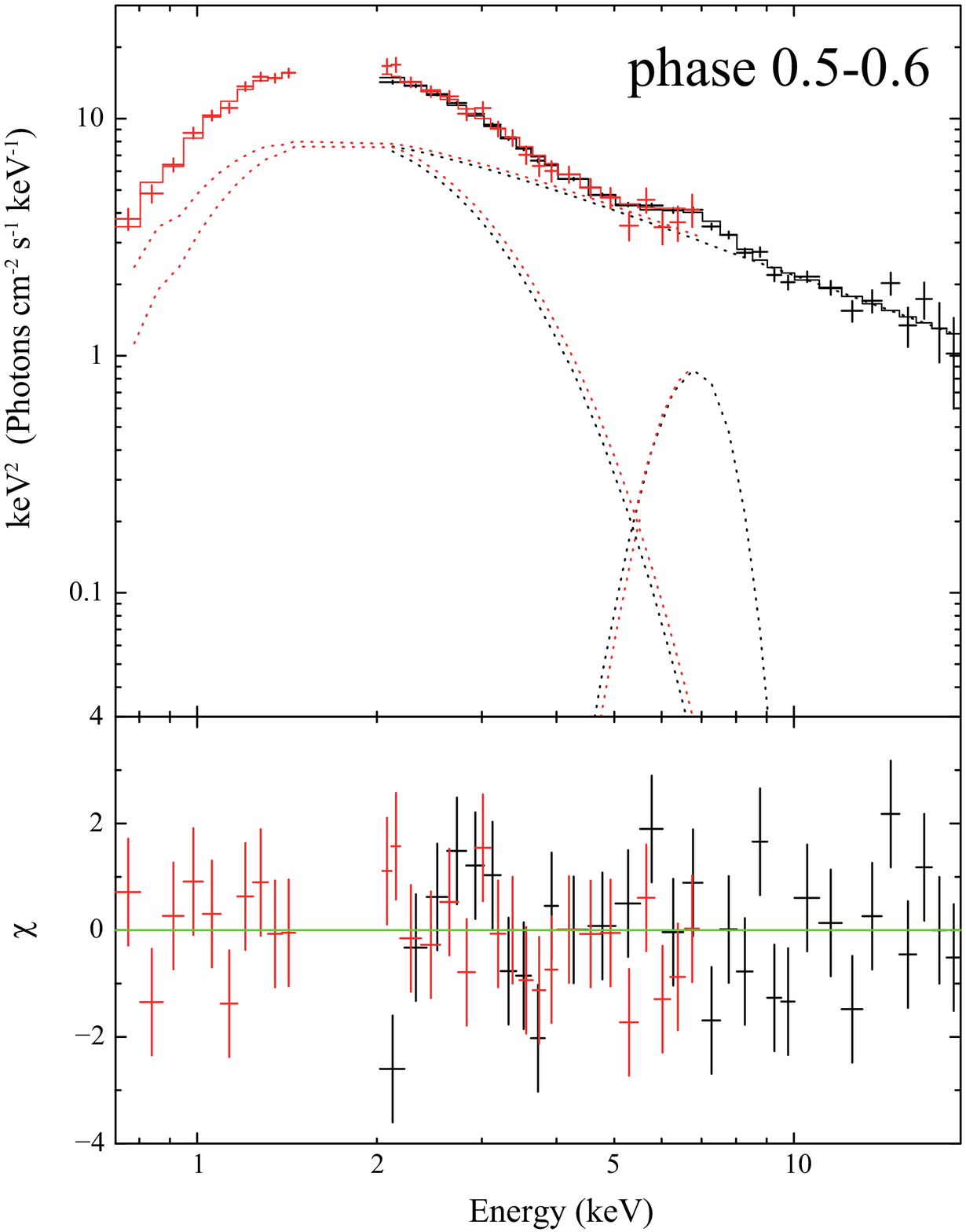}   
  \end{center}
  \caption{
  Background-subtracted unfolded spectra in the $0.0-0.1$ and in the $0.5-0.6$ 
  in the LHS (upper figures) and the HSS (lowers).
  The black and the red are the GSC and the SSC spectra, respectively.
  The model is {\em phabs*powerlaw} for the LHS and 
  {\em phabs*(diskbb+powerlaw+gaussian)} for the HSS.
  The $1.5-2.0$ keV energy range of the SSC spectrum is ignored in the fitting.
  The dotted lines represent contributions of the three components.
  The bottom panels are residuals from the model.
  }
\label{fig:spec_9-1}
\end{figure*}

In order to parameterize the spectral modulation, 
we performed a model fitting of the extracted spectra.
We divided the orbital phase into ten phase intervals 
and extracted ten spectra for each state and for each of the GSC and SSC.
The GSC and SSC spectra cover $2-20$ keV and $0.7-7$ keV bands, respectively.
As examples, 
the spectra in the phase $0.0-0.1$ and $0.5-0.6$ in the LHS and HSS 
are plotted in figure \ref{fig:spec_9-1} 
with their best fit models described below.
The 1.5--2 keV range of the SSC was excluded
to avoid the known systematic uncertainty in the effective area (see Paper I).

\subsubsection{Spectral model fitting result in the LHS}

\begin{figure*}
  \begin{center}
   \includegraphics[width=5cm]{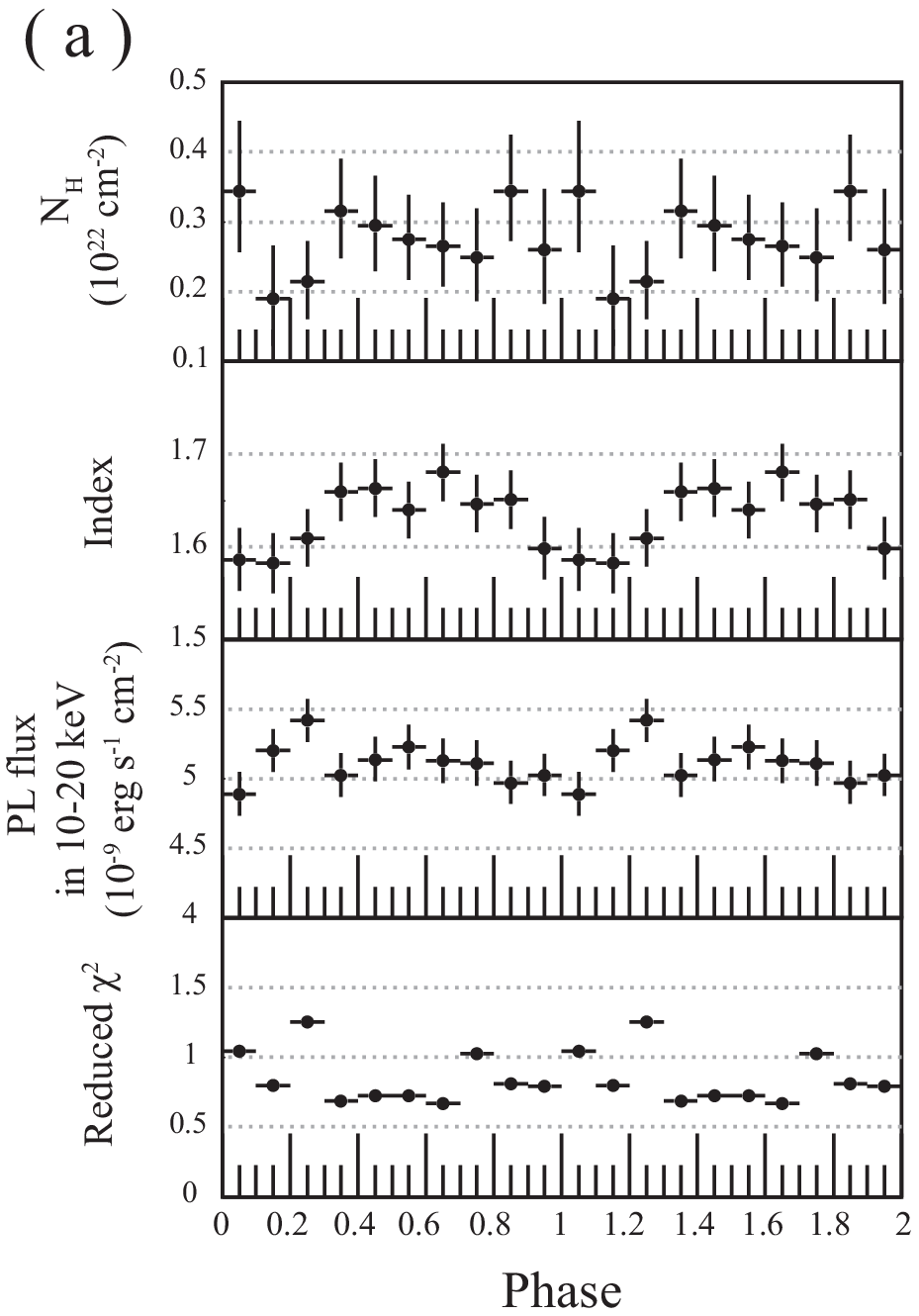}
   \hspace{0.5cm}
   \includegraphics[width=5cm]{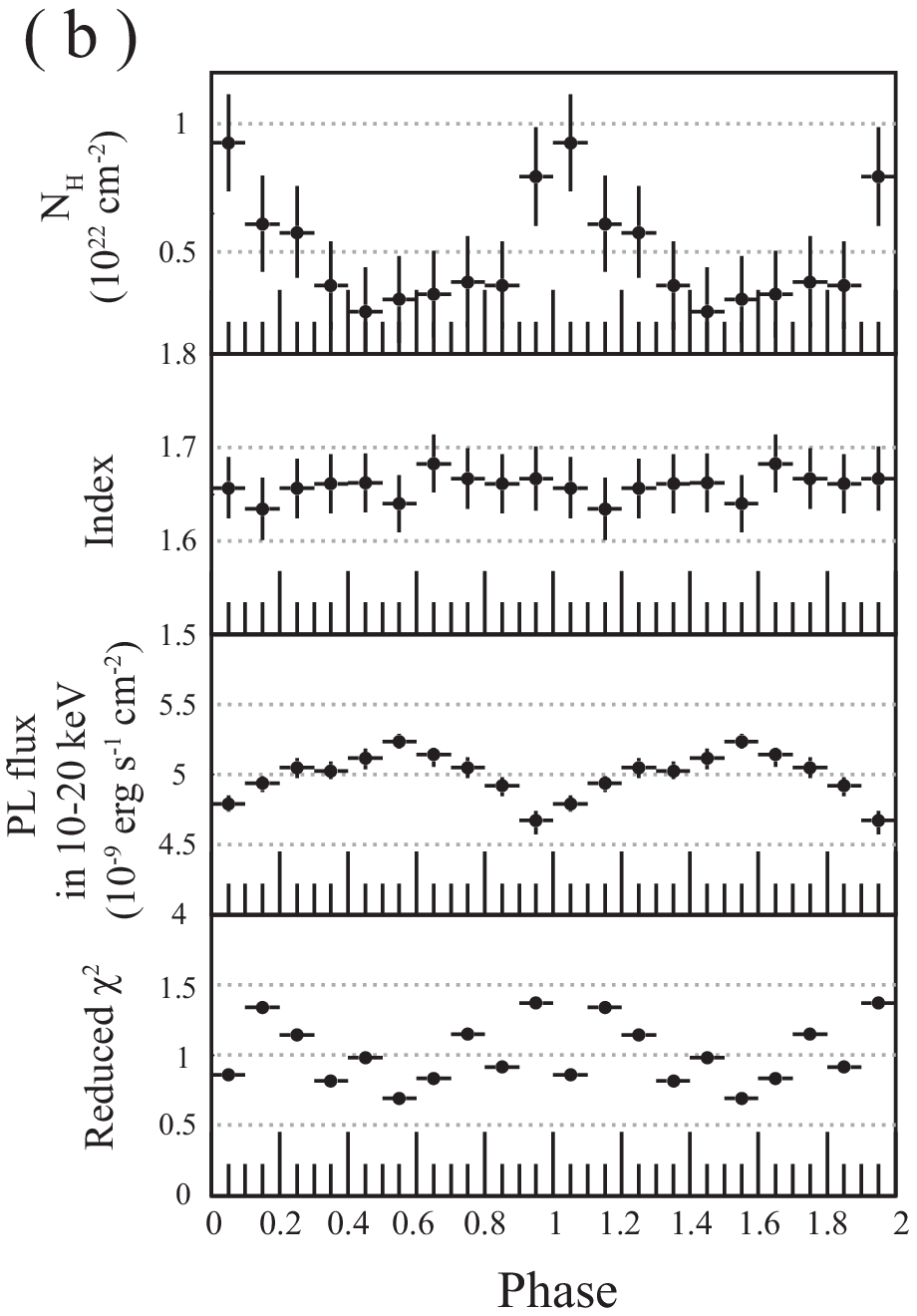}
   \hspace{0.5cm}
   \includegraphics[width=5cm]{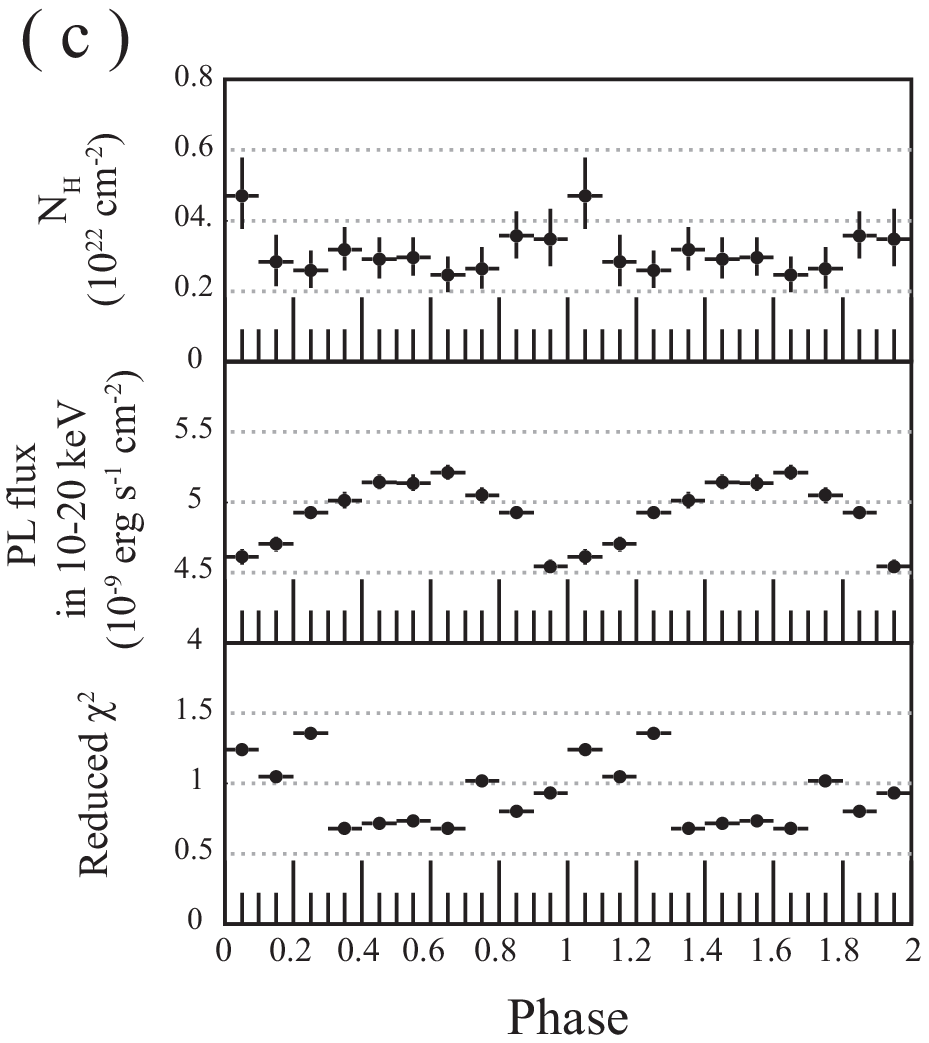}
  \end{center}
  \caption{
  Best fit parameters in the LHS.
  (a) The simultaneous model fitting to the $2-20$ keV GSC spectra and the $0.7-7$ keV SSC spectra.
  The model is {\em phabs*powerlaw}. 
  (b) The model fitting to the GSC spectra only with {\em phabs*powerlaw}. 
  (c) The model fitting to the GSC and the SSC spectra 
  with {\em phabs*powerlaw}. 
  The photon index was fixed at the average value, 1.66.
  }
\label{fig:specanaLHS}
\end{figure*}

First, we fitted a power law ({\em powerlaw}) model 
with a photoelectric absorption ($phabs$) 
to the GSC and the SSC spectra in the LHS. 
We added systematic errors of $1 \%$ to both the GSC and SSC data 
for taking account of uncertainty of the calibration. 
The resultant reduced $\chi^{2}$ values ranged from 0.9 to 2.1, 
and best-fit models in several phase intervals were still not acceptable. 
It was thought that the applied model was not appropriate enough. 
Because an application of more sophisticated model would be beyond the scope of this work, 
we increased the systematic error to $5\%$, so as to include the model uncertainty. 
Then reduced $\chi^{2}$ values became less than 1.3.
The obtained parameters were plotted in figure \ref{fig:specanaLHS} (a). 
As the errors were large, 
we did not recognize a modulation of the hydrogen column density, $N_{\rm H}$, 
and of the {\em powerlaw} flux in $10-20$ keV, 
but we found a modulation of the power law index. 
On the other hand, 
in figure \ref{fig:specanaLHS} (b), 
we plotted the best-fit parameters 
obtained from the fitting of the GSC data only. 
The GSC results indicate that 
the power law index is not modulating within $\pm 0.05$ and 
$N_{\rm H}$ shows a modulation. 
The average value of the power law index is $1.66 \pm 0.02$. 
These facts mean that 
the SSC data shows an modulation 
which can not be well described by a simple absorption model, 
suggesting an partial absorption or an existence of a low temperature disk component. 
However, since the farther investigation of the complex spectral shape is not easy by our data, 
we applied again the model, {\em phabs*powerlaw}, 
with a fixed power law index of 1.66, 
which is the value determined by the GSC.
Therefore 
the complex spectral variation of the low energy part is all approximately expressed 
by a change of $N_{\rm H}$.
The best fit parameters were shown in figure \ref{fig:specanaLHS} (c). 
We found a marginal suggestion of an increase of $N_{\rm H}$ around phase 0 
with $\Delta N_{\rm H} = (1.8 \pm 1.2) \times 10^{21}$ cm$^{-2}$ 
(from $(2.9 \pm 0.4) \times 10^{21}$ cm$^{-2}$ to $(4.7 \pm 1.1) \times 10^{21}$ cm$^{-2}$). 
We also found that 
the {\em powerlaw} flux, in the 10-20 keV band, had a modulation with $MF = 6 \pm 1 \%$. 
The large reduced $\chi^{2}$ values around phase 0 suggest that 
spectra around phase 0 was not simulated well 
by the simple power law model and 
may need more complex model, such as a partial absorption model, 
due to averaging of many spectra with various $N_{\rm H}$ values. 
The value of $N_{\rm H}$ obtained from our analysis is smaller 
than those in previous works 
by \citet{kitamoto1984} ($N_{\rm H} =  5 \times 10^{21} \sim 2 \times 10^{22}$ cm$^{-2}$) and 
\citet{grinberg} ($N_{\rm H} = 1 \times 10^{22} \sim 2 \times 10^{22}$ cm$^{-2}$).
Since our data covers down to 0.7 keV, 
the possible partial absorption may affect the resultant $N_{\rm H}$ values to be small with more sensitivity 
than the above previous reports. 
This is supported by the figure \ref{fig:specanaLHS} (b), 
where the $N_{\rm H}$ around phase 0, 
obtained with the GSC only,  is $\sim 1 \times 10^{22}$ cm$^{-2}$.

\subsubsection{Spectral model fitting result in the HSS}

\begin{figure*}
  \begin{center}
   \includegraphics[width=5cm]{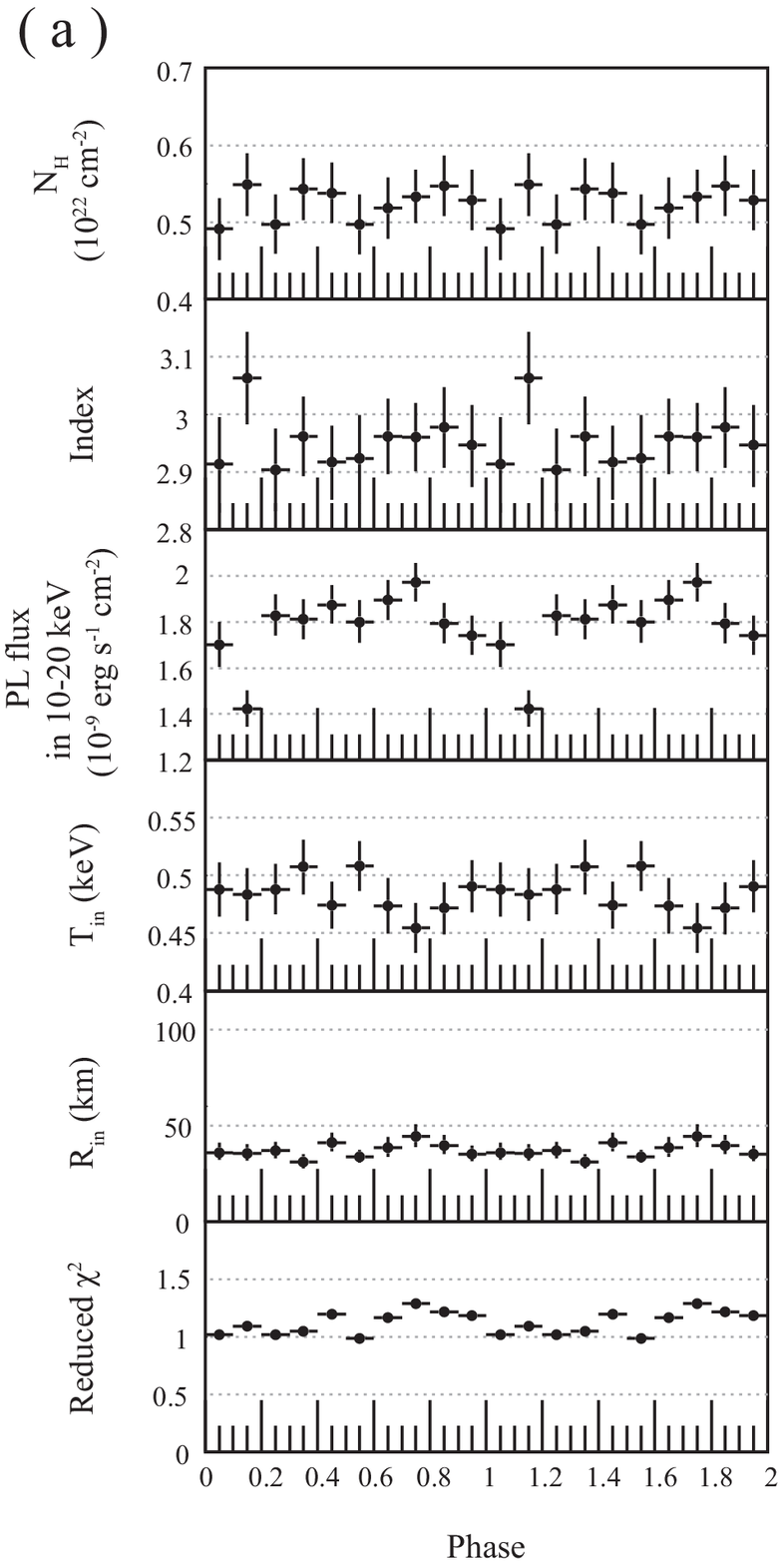}
   \hspace{0.5cm}
   \includegraphics[width=5cm]{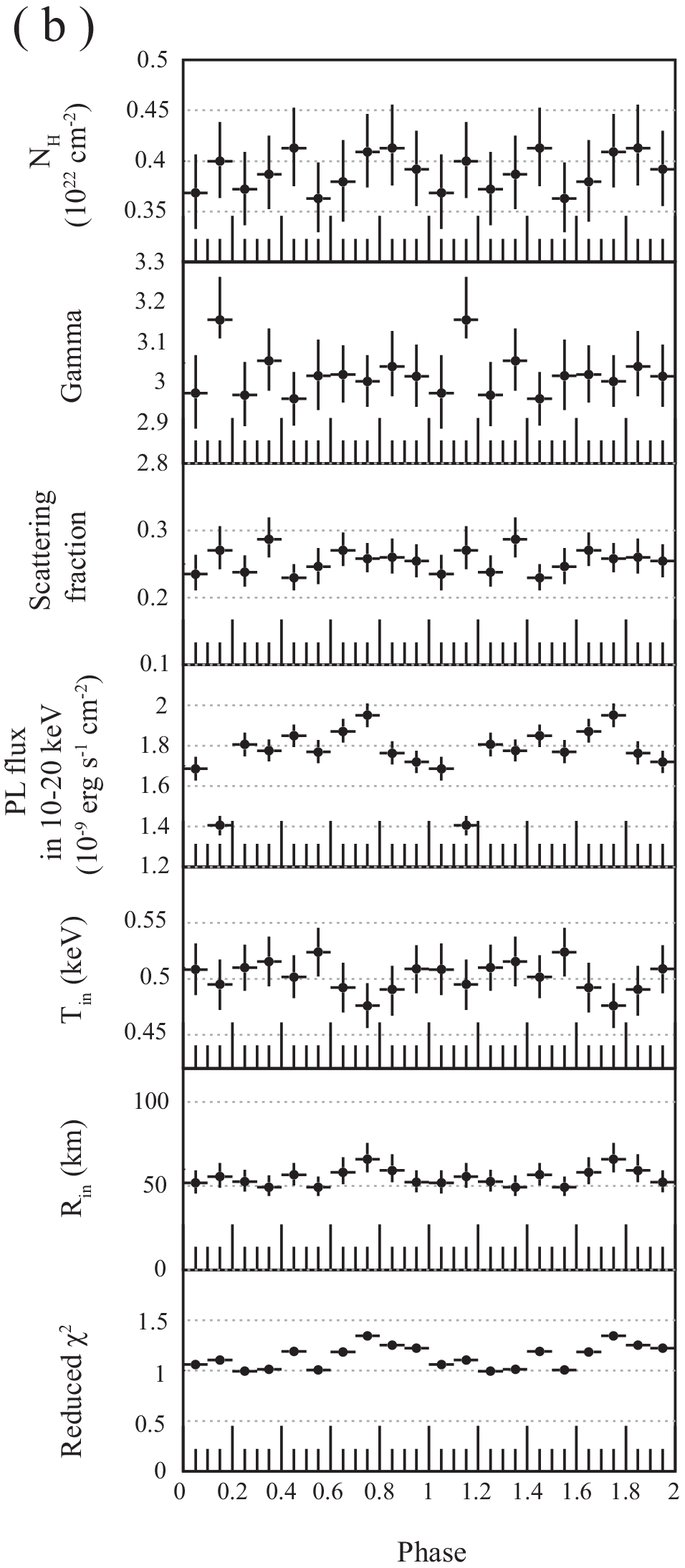}  
  \end{center}
  \caption{
  Best fit parameters in the HSS.
  (a) The simultaneous model fitting to the GSC and the SSC spectra.
  The model is {\em phabs*(powerlaw+diskbb+gaussian)}.
  (b) The model fitting to the GSC and the SSC spectra
  with {\em phabs*simpl*(diskbb+gaussian)}.
  }
\label{fig:specanaHSS}
\end{figure*}

The spectra in the HSS were first fitted by a model, {\em phabs*(powerlaw+diskbb+gausian)},  
where {\em diskbb} is a multi-color disk model \citep{mitsuda}
and 
{\em gaussian} is for a Fe-K emission line.	
As with the LHS, we added 1\% systematic errors.  
The reduced $\chi^{2}$ values were all less than 1.3, indicating reasonable fittings.
Best fit parameters are shown in figure \ref{fig:specanaHSS} (a).
In the HSS, 
$N_{\rm H}$ did not increase around phase 0 
and the variation was not more than $1\times 10^{21}\ {\rm cm}^{-2}$.
The innermost radius and temperature of the disk component were $\sim 40$ km and $\sim 0.5$ keV, 
where we assumed the distance of 1.86~kpc and the inclination of 27$^{\circ}$ \citep{orosz}.  
We did not find  modulation of them. 
The modulation of the {\em powerlaw} flux was $7\pm 5 \%$ in $MF$.
Here, we should note that 
the power law index is $\sim 3$. 
Such a steep {\em powarlaw} model affects the spectra in the low energy band, 
where it is not realistic. 
Then,  
we substituted a ''{\em simpl}" model, ({\em phabs*simpl*(diskbb+gaussian)}), 
which is an empirical model of Comptonization : 
a fraction of photons in an input seed spectrum is scattered into a power law component \citep{steiner}.
The best fit parameters are plotted in figure \ref{fig:specanaHSS} (b).
The change of $N_{\rm H}$ was again less than $1\times 10^{21}\ {\rm cm}^{-2}$, 
and the absolute values became small.
The modulation of the flux in the $10-20$ keV was again $7 \pm 3 \%$ in $MF$. 
So far the orbital modulation of $N_{\rm H}$ in the HSS was only reported by \citet{grinberg}. 
They reported that 
their data was still poor and 
their derived values of $N_{\rm H}$ had large scatting 
with $2 \sim 3 \times 10^{22}$ cm$^{-2}$ on the average 
and $0 \sim 3 \times 10^{22}$ cm$^{-2}$ on  the median, 
and no clear orbital modulation was reported.

\section{Discussion}
\subsection{Orbital Modulation}

We first clearly detected the orbital modulation of the X-ray intensity in the HSS, 
which was an energy independent modulation with $MF$ of roughly $4 \%$ up to 10 keV.
In the $10-20$ keV band, data was not in-consistent with the same amount of the modulation.
On the other hand, in the LHS, 
an enhanced low energy modulation was notable below 4 keV, 
overlapping to the energy-independent modulation with $MF$ of $\sim 4 \%$. 
A reasonable interpretation is that 
the stellar wind is almost ionized and electron scatterings by the ionized wind make the energy independent modulation.
In the LHS, 
the wind in the line of sight at the superior conjunction of the BH is not ionized enough 
and the photoelectric absorption is observed.
Our data showed the increase of $N_{\rm H}$ around the phase 0 in the LHS as 
$(1.8 \pm 1.2) \times 10^{21}\ {\rm cm}^{-2}$ 
for fitting with {\em powlaw} model 
on the condition that 
the parameters of the photon index was fixed.
This value was, however,  smaller than the previous reports of the absorption dips, 
for example, $10^{23}\ {\rm cm}^{-2}$ as reported by \citet{grinberg} and \citet{kitamoto1984}.
It is also known that 
the dip depth and the dip duration have wide variety \citep{church} 
and the dip does not always occur in the superior conjunction \citep{kitamoto1984}.
The parameter obtained by our MAXI observation should be recognized to be a kind of averaged value of $N_{\rm H}$, 
among various snapshot-samplings around the superior conjunction of the BH 
with variety of high density region in their size, density and 
even in their possible ionization state \citep{feng}.

\citet{yamada} reported a detection of a dip during the HSS of Cyg X-1, 
which showed a spectral hardening.
They reported the detection of the absorption lines of highly ionized irons around the phase 0, 
and its equivalent width became large during the dip. 
This indicates that 
the wind is highly ionized in the HSS,
but some blob-like structure with high density may  still enhance the low energy absorption in a short time, 
and make the ''so-called" absorption dip with a spectral hardening.
Since our data did not show the increase of the $N_{\rm H}$ around phase 0 in the HSS,  
we should consider that the ''so-called" absorption dip is rare in the HSS.

\subsection{Density of stellar wind}

We examined 
whether our finding of 
the roughly sinusoidal and energy-independent modulation 
can be explained by a reasonable wind-parameters or not.  
Although an actual wind has a complex structure such as the focusing wind 
(e.g. \cite{misko}), 
in this work 
we concentrate only the overall sinusoidal modulation with the $3 \sim 4 \%$ amplitude and 
assume a much simple toy model.  
We used the wind parameters reported by \citet{vrtilek}, 
and the inclination reported by \citet{orosz}, 
listed in table \ref{tb:para}.
We assumed that 
the wind is spherically symmetric, 
and thus  
the wind density, $n(r)$, 
at the distance, $r$, from the companion star 
can be expressed as
\begin{equation} 
n(r) = \frac{\dot{M}}{4\pi m_{\rm H}v(r)}\frac{1}{r^2}
\label{eq:dens}
\end{equation}
where $\dot{M}$ is the mass loss rate as the stellar wind, 
$m_{\rm H}$ is the mass of the hydrogen atom 
and $v(r)$ is the wind velocity.
If we express the ionization fraction of Hydrogen as $\kappa$,  
the electron column density, $N_{\rm e}$, can be calculated as
\begin{equation} 
N_{\rm e} = \int_{0}^{\infty}\kappa(l) n(l) dl
\label{eq:dens}
 \end{equation}
where the integral is performed along the line of sight, $dl$, from the BH to the observer.
If we approximate the wind velocity as a constant value with the terminal velocity, $v_{\infty}$,  
and the ionization fraction $\kappa(l)$ is constant as $\kappa_0$, 
the integral can be reduced as 
\begin{eqnarray} 
N_{\rm e} & = & \frac{\dot{M} \kappa_0}{4\pi m_{\rm H}v_{\infty}}\frac{\pi -\chi}{a\sin \chi},\\
\cos{\chi} & = & \sin i \cos \theta \nonumber
\label{eq:nh}
 \end{eqnarray}
where $\chi$ is the angle between the line of sight and the line joining the centers of two stars, 
represented as $\cos^{-1}(\sin i\cos \theta)$, 
$i$ is the inclination angle and $\theta$ is the orbital phase.
Since the wind condition may change between two states, 
we introduce a parameter $\eta = \kappa_0 \frac{\dot{M}/\dot{M}_0}{v_{\infty}/v_{\infty,0}}$, 
where $\dot{M}_0$ and $v_{\infty, 0}$ are the values as listed in table \ref{tb:para}.

\begin{table*}
\caption{Parameters used in calculation.}
\begin{center}
\begin{tabular}[b]{c c c}
\hline
Parameter    & Value  & Reference \\ 
\hline
Mass loss rate~($\dot{M}_0$)         & $\sim 5\times 10^{-6}\ M_{\odot}$ yr$^{-1}$  & \citep{vrtilek}  \\
Terminal velocity~($v_{\infty,0}$)   & $\sim$1400\ km s$^{-1}$                      & \citep{vrtilek}\\
$\beta$ of the CAK$^*$ wind model    & $\sim$0.75                                   & \citep{vrtilek}\\
Stellar radius~($R_{*}$)             & $\sim1.5\times 10^{12}$ cm                   & \citep{vrtilek} \\
Separation of the binary~($a$)       & $\sim 3.0\times 10^{12}$ cm                  & \citep{vrtilek} \\
Inclination of the binary~($i$)      & 27.1$^{\circ}$                               & \citep{orosz}\\
\hline
\end{tabular}
\\$^*$~\citet{CAK}
\end{center}
\label{tb:para}
\end{table*}

\begin{figure*}
  \begin{center}
   \includegraphics[width=5cm]{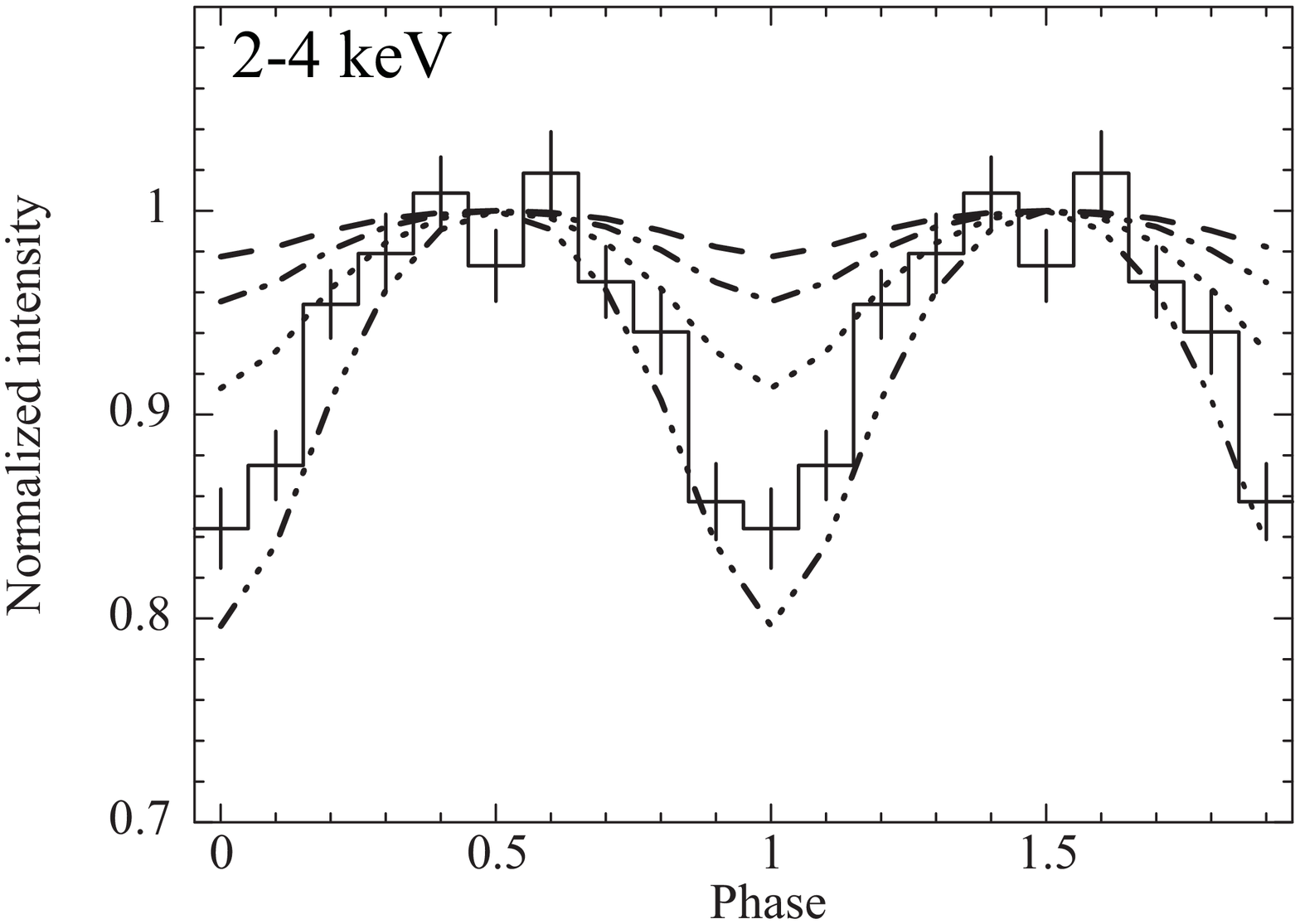}
   \includegraphics[width=5cm]{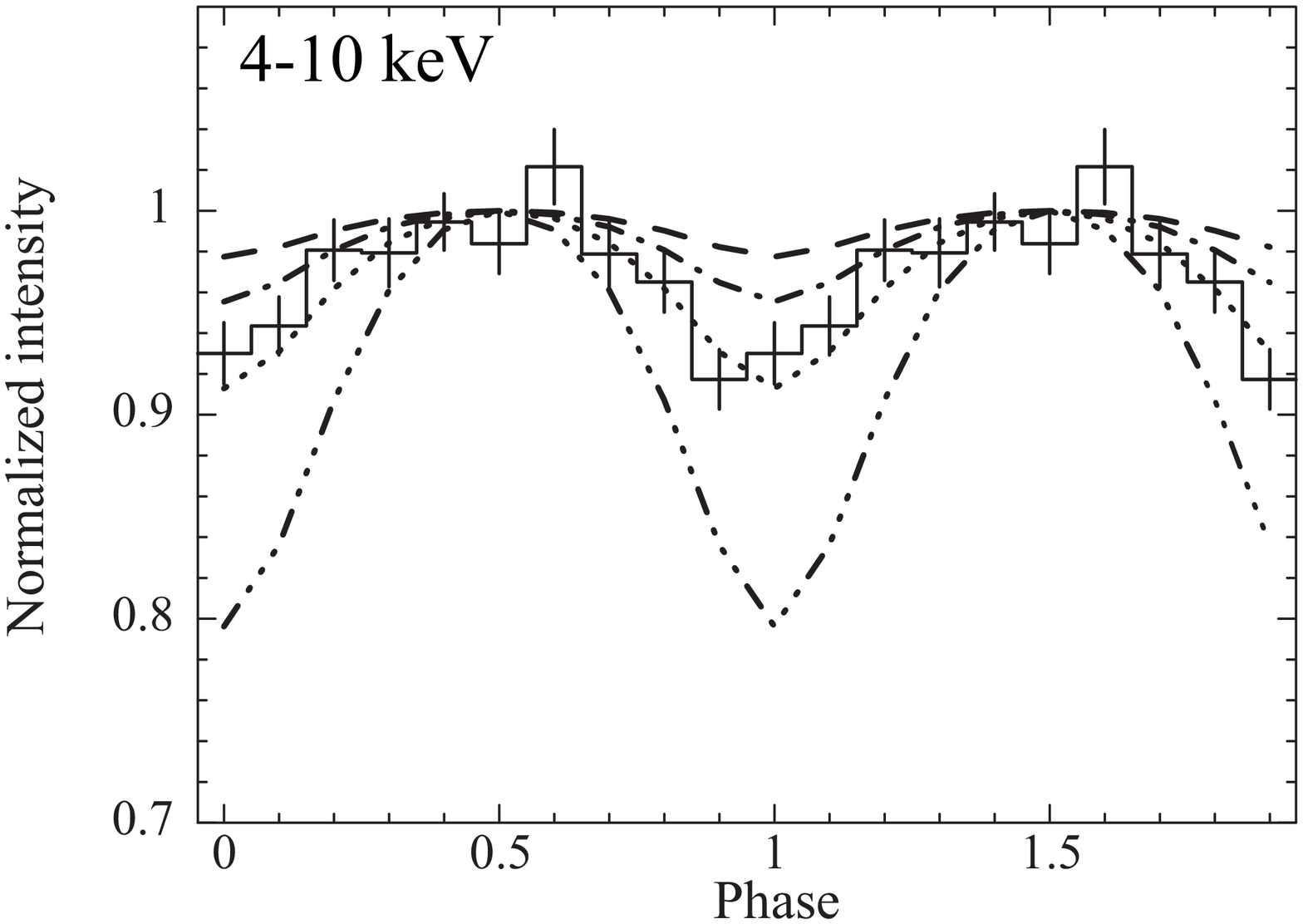}
   \includegraphics[width=5cm]{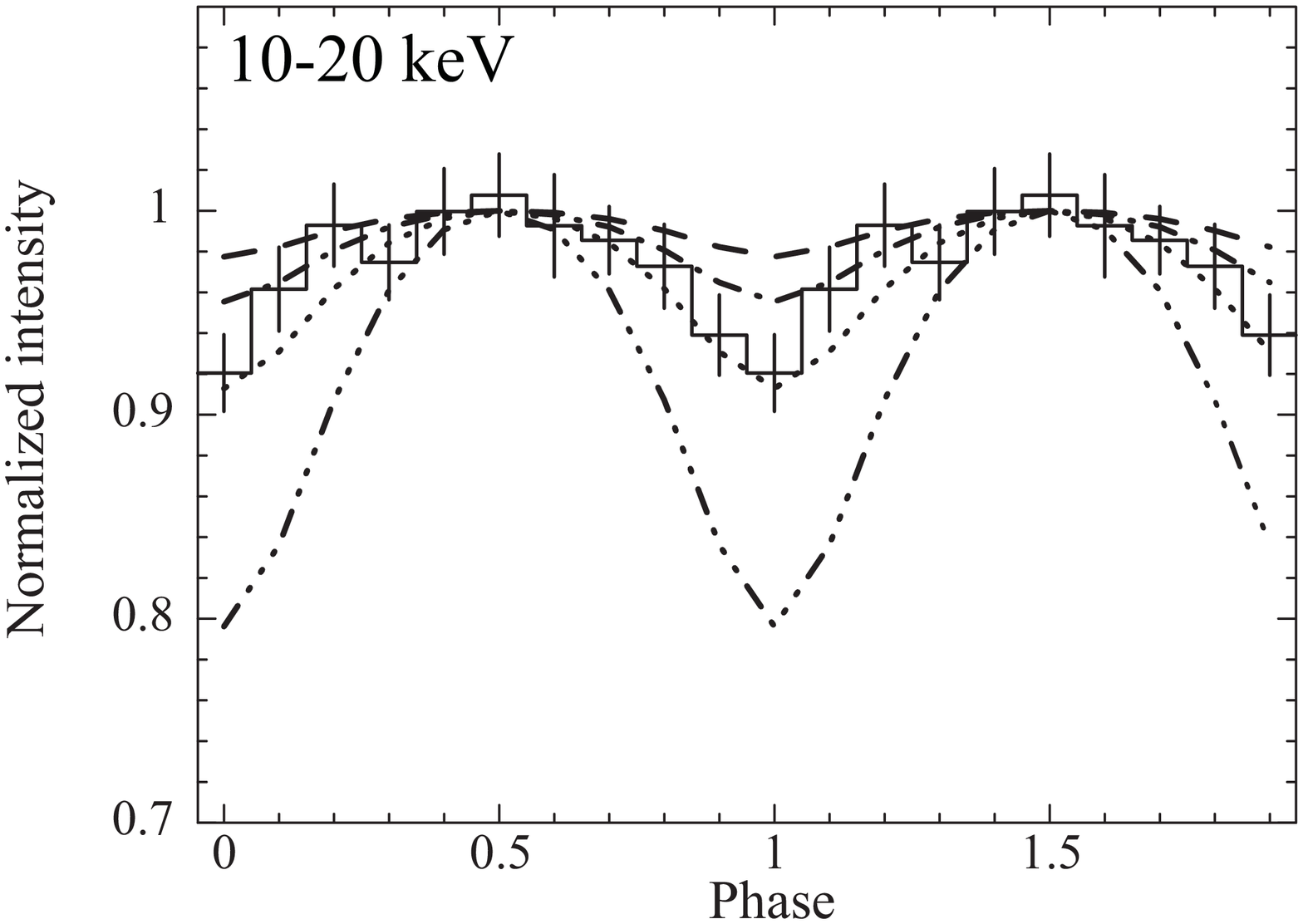}\\
   \includegraphics[width=5cm]{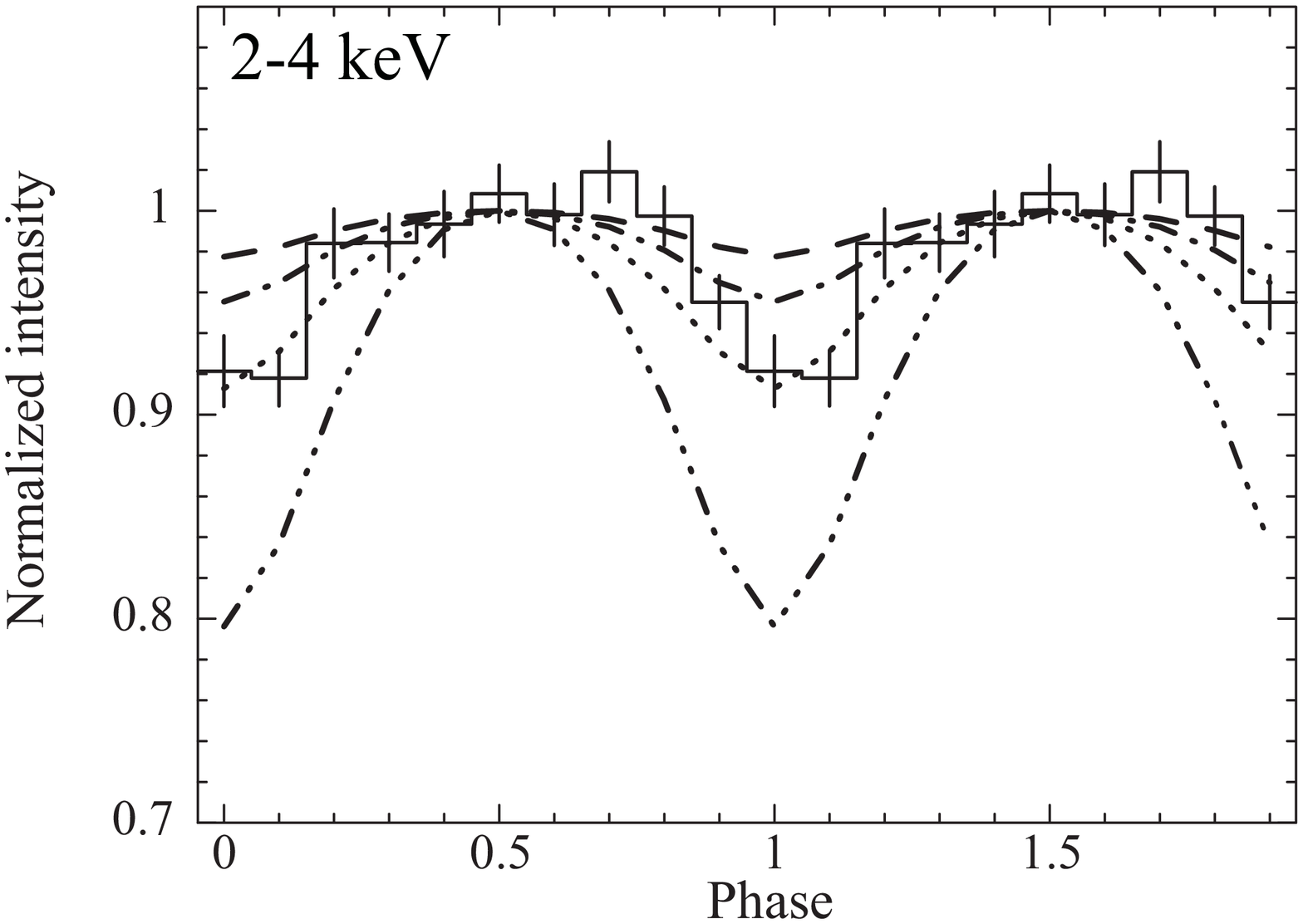}
   \includegraphics[width=5cm]{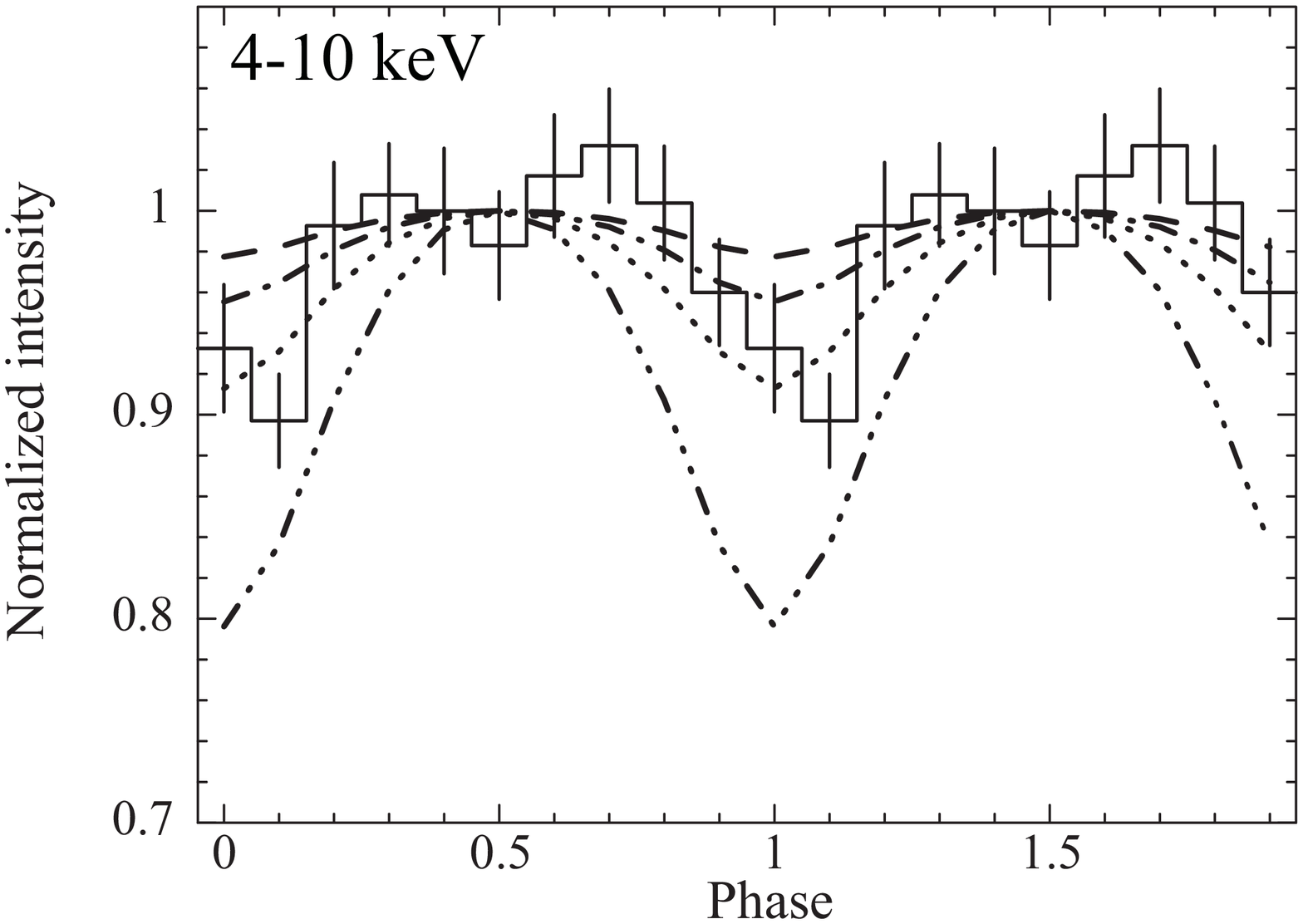}   
   \includegraphics[width=5cm]{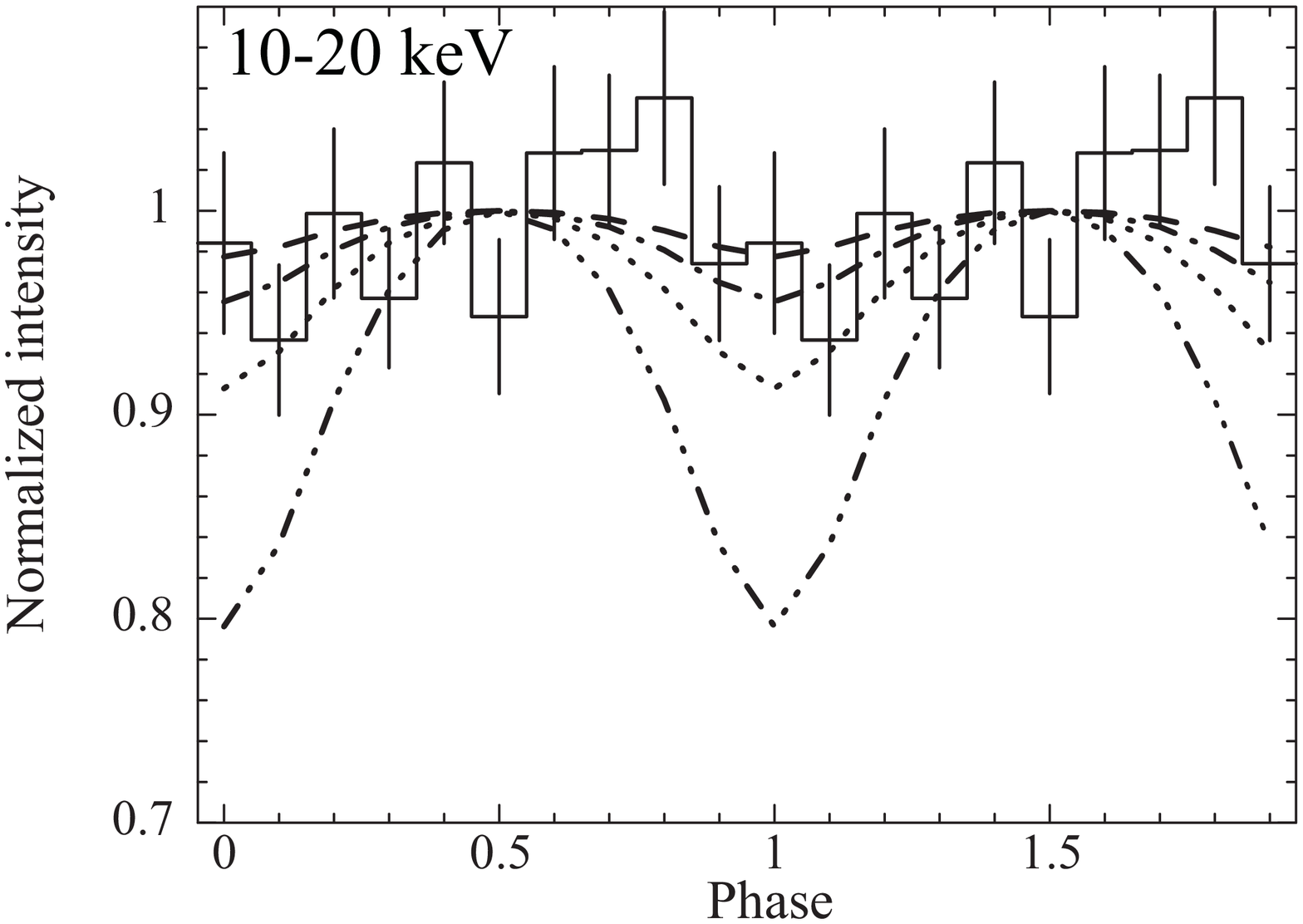}   
  \end{center}
  \caption{The expected light curves for four values of parameter $\eta$ 
  in the LHS (upper panels) and in the HSS (lower panels).
  Dashed, dashed-dotted, dotted and dashed-dotted-dotted lines are data points with $\eta = 0.5, 1, 2, 5$, respectively.
  The observed folded light curves are also potted with solid lines. 
  All the light curves are adjusted to be one at the peak intensity.}
\label{fig:lcsim}
\end{figure*}

Using the parameters listed in table \ref{tb:para} and the Thomson scattering cross section, $\sigma_{\rm T}$, 
we can calculate an optical depth for the electron scatting.
The intensity variation can be roughly expressed as 
$I =I_0 \exp{(-N_{\rm e} \sigma_{\rm T})}$, 
as long as the optical depth is small, 
where $I_0$ is the original intensity from near the BH.
In figure \ref{fig:lcsim}, 
the expected light curves for the several cases (0.5, 1, 2, 5) of the parameter $\eta$ are plotted 
as well as the observed folded light curves.
All the curves were adjusted to be one at the peak intensity.

Except for the light curve in the energy range $2-4$ keV around phase 0 in the LHS, 
the light curves were able to be simulated with the parameter $\eta = 2.0$ 
($N_{\rm{e}} = (0.8-1.6) \times 10^{23}\ \rm{cm}^{-2}$), 
in both states.
This suggests that 
the stellar wind parameter, $\eta$,  
does not change between the two states within the precision of our observation.
The notable discrepancy was seen 
around the phase 0 in the low energy band of the LHS. 
This is interpreted that 
the amount of the photoelectric absorption increases 
but that of electron scattering is not. 
Therefore, 
just on the line of sight around the phase 0, 
the ionization state of the stellar wind is low in the LHS 
comparing to that of HSS.
\citet{grinberg} calculated the wind column density of $3 \sim 5 \times 10^{22}$ cm$^{-2}$, 
using more realistic model assuming the focusing wind, 
CAK wind model and parameters reported by \citet{gies}. 
Although our model is a simple toy model, the obtained parameter, $N_{\rm e} \sim 10^{23}$ cm$^{-2}$, 
requires two of three times higher values than the above model parameters, 
i.e. slow velocity or high mass loss rate.	

\subsection{Stellar wind model}

The ionization should be due to the photoionization by intense X-rays from near the BH.
In such a case, the ionization state can be estimated 
by the ionization parameter, $\xi = \frac{L_x}{n d^2}$, 
where $L_x$ is the X-ray luminosity of the compact source, 
$n$ is the ion density 
and $d$ is the distance from the compact source \citep{tarner}.
According to the calculation of the model 6 and 7 by \citet{kallman}, 
an ionization front of the He can be seen around $\xi \sim 30$.
If $\xi$ exceeds 30, 
the majority of the metal is ionized, 
but if $\xi$ is less than 30, 
substantial amount of the metal is not ionized and 
the absorption by metal becomes notable.

From the fitting result of the energy spectrum, 
we obtained that the luminosity in $0.7-7$ keV was $0.4\times 10^{37}$ erg s$^{-1}$  
and  $2.4\times 10^{37}$ erg s$^{-1}$ 
for the LHS and HSS, respectively.
This six times difference of the luminosity should make the difference of the ionization state of the wind  
between the two states.
On the other hand, 
we considered that 
the energy independent intensity modulation is caused by the electron scattering 
and the parameter $\eta$ is about 2 for the both states.
Except for phase 0 of the LHS, we did not detect any orbital modulation of the amount of the photo-electric absorption 
in spite of non-spherical-symmetry of the wind column density seen from the BH.  
Therefore, we can assume, as a first approximation, that the ionization state of the wind is enough high and 
the un-known parameter $\kappa_0 =1$. 
Then we can examine the ionization parameter at the nearest point of the companion star along the line of sight at the phase 0, 
using the parameters listed in table 1 and the $\eta$ of 2, as 

\begin{equation} 
\xi  \sim 500 L_{x, 38} 
\label{eq:dens}
\end{equation}
where $L_{x,38}$ is the X-ray luminosity with an unit of 10$^{38}$ erg s$^{-1}$. 
Therefore the wind is ionized well at the point in the HSS ($L_{x,38} \sim 0.24$) .
In the LHS ($L_{x,38} \sim 0.04$), 
however, a delicate balance between the luminosity and the density leads large variation of the ionization state around the phase 0.
Various high density blobs, in an in-homogeneous wind work 
as photoelectric absorbers around the superior conjunction of the BH, 
and cause the various dips in the LHS.
In the other phase, 
most of the wind in the LHS, as well as the HSS, is ionized, 
and this is consistent to the high resolution observation in the LHS \citep{misko}.

\section{Conclusion}

We analyzed the orbital variability of Cyg X-1 in the LHS and the HSS with MAXI observation.
We detected, for the first time,  
an intensity modulation with the orbital period in the HSS, 
as well as that in the LHS.
The modulation in both states can be explained 
by the electron scattering of the ionized stellar wind of the companion star, 
except for the phase around the superior conjunction of the BH in the LHS, 
where additional photoelectric absorption is required.
These condition can be explained by  reasonable  parameters of an in-homogeneous wind 
and by the luminosity of the states.

\bigskip

This work was supported by RIKEN Junior Research Associate Program. 
This work was also partially supported by the Ministry of Education, Culture, Sports, Science and Technology (MEXT), 
Grant-in-Aid for Science Research 24340041, 
and the MEXT Supported Program for the Strategic Research Foundation at Private Universities, 2014-2018.

\end{document}